\documentclass[lineno]{biometrika}

\usepackage{newtxtext}
\usepackage[subscriptcorrection]{newtxmath}
\usepackage{graphicx, amsmath, bm, mathtools, comment, environ}

\newcommand{\elemabs}[1]{(#1)^{\mathrm{abs}}}

\newenvironment{modnew}{\color{black}}{\unskip}

\DeclareMathOperator{\diag}{diag}
\DeclareMathOperator{\re}{Re}
\DeclareMathOperator{\im}{Im}

\begin{document}

\jname{Biometrika}


\markboth{Yiran Zeng \and Dale L. Zimmerman}{Characteristic function-based tests for spatial randomness}

\title{Characteristic function-based tests for spatial randomness}

\author{Yiran Zeng}
\affil{Department of Statistics and Actuarial Science, University of Iowa, Iowa City, Iowa 52242,
U.S.A.
\email{yiran-zeng@uiowa.edu}}

\author{\and Dale L. Zimmerman}
\affil{Department of Statistics and Actuarial Science, University of Iowa, Iowa City, Iowa 52242,
U.S.A. \email{dale-zimmerman@uiowa.edu}}

\maketitle

\begin{abstract}
We introduce a new type of test for complete spatial randomness that applies to mapped point patterns in a rectangle or a cube of any dimension. This is the first test of its kind to be based on characteristic functions and utilizes a weighted $L_2$-distance between the empirical and uniform characteristic functions. The test shows surprising connections to Ripley's $K$-function and Zimmerman's $\bar{\omega}^2$ statistic. It is also simple to calculate and does not require adjusting for edge effects. An efficient algorithm is developed to find the asymptotic null distribution of the test statistic under the Cauchy weight function. This makes the test fast to compute.
In simulations, our test shows varying sensitivity to different levels of spatial interaction depending on the scale parameter of the Cauchy weight function. Tests with different parameter values can be combined to create a Bonferroni-corrected omnibus test, which is more powerful than the popular $L$-test and the Clark-Evans test in most simulation settings of heterogeneity, aggregation and regularity, especially when the sample size is large. The simplicity of the empirical characteristic function makes it straightforward to extend our test to non-rectangular or sparsely sampled point patterns.
\end{abstract}


\section{Introduction}

Tests for complete spatial randomness (CSR) play an important role in the analysis of spatial point patterns. The outcome of such a test determines whether a model more complicated than the simple Poisson process is necessary \citep{illian2007intro}. Due to its importance, an extensive literature has been dedicated to developing  CSR testing procedures.  \begin{modnew}
    Within it
\end{modnew}, \citet{illian2007intro} recommends the $L$-test \citep{ripley1979} for fully mapped patterns (the case where the location of every point is available) and the Clark-Evans test \citep{clarkevans} for most other cases. In this paper, we restrict our attention to fully mapped patterns, although both  \begin{modnew}
    the $L$-test and Clark-Evans test
\end{modnew} will be included in the numerical analysis. 

First described in \citet{Ripley1978}, the $L$-test measures how much the $L$-function of a given pattern differs from that  \begin{modnew}
    expected under
\end{modnew} CSR . The $L$-function is defined in terms of Ripley's $K$-function, $L(r) = \{K(r) / \pi\}^{1/2}$, where $K(r)$ is the expected number of random points within a distance $r$ of  \begin{modnew}
    an arbitary
\end{modnew} point, divided by the intensity. The $L$-function has a very simple form, $L(r)=r$,  \begin{modnew}
    under CSR
\end{modnew}. The $L$-test statistic $L_m$ is then the maximum absolute difference between the estimated $\hat{L}(r)=\{\hat{K}(r) / \pi\}^{1/2}$ and its theoretical value $r$ under the null hypothesis, i.e., $L_m = \sup_{r \le s} |\hat{L}(r) - r|$. Large $L_m$ values  lead to the rejection of the null hypothesis. The purpose of the upper bound $s$ is to nullify high fluctuations of $\hat{L}(r)$ at large distances. \citet{Ripley1988} provides an estimator for $K(r)$ with isotropic edge correction.

The much older Clark-Evans test statistic \citep{clarkevans} is a  \begin{modnew}
    function of
\end{modnew} the average nearest-neighbor distance. Small Clark-Evans statistics indicate aggregation and large values indicate regularity. The Clark-Evans test is more versatile in that it does not require the data to be fully mapped, but  it is not very powerful against regular patterns \citep{illian2007intro}.

 \begin{modnew}
    Conditional on the number of points, a
\end{modnew} CSR point process can also be viewed as a collection of independent uniformly distributed random variables, so the classical discrepancy (or goodness-of-fit) statistics, such as the Kolmogorov-Smirnov, Cram\'er-von Mises, and Anderson-Darling statistics, are also suitable for testing for CSR, especially when the sampling window is rectangular \citep{zimmerman1993}. They are measures of distance between the empirical cumulative distribution function (CDF) and the theoretical CDF of a two- (or more) dimensional uniform distribution. \citet{Ho2007} notes that  discrepancy tests have the advantage of not requiring edge corrections, user-chosen parameters, or the estimation of intensity. \citet{zimmerman1993}'s $\bar{\omega}^2$ is a test for CSR of Cram\'er-von Mises type, modified to be independent of the choice of origin. Large values of $\bar{\omega}^2$ indicate aggregation or heterogeneity and small values indicate regularity. The asymptotic null distribution for $\bar{\omega}^2$  \begin{modnew}
    was
\end{modnew} derived by \citet{ZIMMERMAN1994189}, allowing for easy testing. This test is particularly strong against heterogeneous alternatives, i.e., global deviations from uniformity. \citet{Ho2007}  \begin{modnew}
    have considered
\end{modnew} other types of discrepancy tests that are also strong against long-range clustering. However, tests based on the empirical CDF are not very successful against regularity and short-range clustering.

The characteristic function (CF) can also be used to develop discrepancy tests. It also does not require edge corrections or the estimation of intensity, but unlike the empirical CDF, the empirical CF can capture high-frequency features in the distribution \citep{eubank1992, FAN1997}, which under a certain interpretation include short-range clustering in the context of a spatial point process, although their connection with regularity is more difficult to qualify theoretically. The sensitivity to high-frequency features could be attributed to the fact that the CF is the Fourier transform of the distribution's density into the frequency domain. Additionally, the empirical CF has a simple unified formula regardless of what the sampling region looks like\textemdash a disk, a polygon, or even a disconnected (or sparsely sampled) region\textemdash and thus is much easier to calculate than the empirical CDF in these scenarios, although this will not be the main focus of this paper. Therefore, CF-based discrepancy tests may be good candidates for CSR testing. However, in the context of spatial point processes, the only paper that deals with CF-like quantities is \citet{rajala}, where they investigate the Fourier transform of the second moment measure of a homogeneous point process. In this paper, we  develop  a CF-based test for CSR. 

CF-based discrepancy tests appear frequently in the broader statistical literature, such as in \citet{koutrouvelis1980}, \citet{koutrouvelis1981}, and \citet{FAN1997}. The particular form of the CF test that we intend to use is the one developed by \citet{epps1983} and \citet{baringhaus1988} for testing normality, and generalized to other null distributions by \citet{epps2005} and \citet{JIMENEZGAMERO20093957}. This test statistic is the weighted $L_2$-distance between the empirical and theoretical CF.
\begin{modnew}
    We analyze this CF test statistic for several types of weight functions. When a Bessel-like weight function is used, we show a surprising similarity between the CF test statistic and Ripley's $K$-function.
\end{modnew}
Remarkably, we \begin{modnew}
    also
\end{modnew} find that \begin{modnew}
    the CDF-based
\end{modnew} Zimmerman's $\bar{\omega}^2$-test is a special case of the CF test for a particular weight function.  The CF test statistic can be plotted against  \begin{modnew}
     a parameter $r$
 \end{modnew} to examine the scale at which  \begin{modnew}
     any
 \end{modnew} deviation from CSR occurs\begin{modnew}
     , just like Ripley's $K(r)$
 \end{modnew}. We also develop a Bonferroni-corrected omnibus test using multiple values of  \begin{modnew}
     parameter $r$
 \end{modnew}. 
 
  Algorithms for the large\begin{modnew}
     -
 \end{modnew}sample null distribution of the test statistic based on the procedure in \citet{JIMENEZGAMERO20093957} and a small\begin{modnew}
     -
 \end{modnew}sample distribution for  \begin{modnew}
     small $r$
 \end{modnew} are  \begin{modnew}
     developed for the Cauchy-weighted CF test statistic
 \end{modnew}. The appropriateness of the null distribution is validated by simulation \begin{modnew}
     in both two and three dimensions
 \end{modnew}. \begin{modnew}
     With the help of the asymptotic null distribution, the Cauchy-weighted CF test is shown to have a much lower computational cost than the isotropic-edge-corrected $L$-test, so we choose it as the focus of our investigation.
 \end{modnew} The \begin{modnew}
     Cauchy-weighted
 \end{modnew} CF test is compared against the $L$-test, Clark-Evans test and Zimmerman's $\bar{\omega}^2$-test in simulation\begin{modnew}
     s
 \end{modnew} and applications. The CF test with parameter  \begin{modnew}
     $r = 1$
 \end{modnew} shows almost no difference in power to the $\bar{\omega}^2$-test. The Bonferroni-corrected omnibus CF test is demonstrated to be more well rounded than the individual CF tests. It beats the Clark-Evans test \begin{modnew}
     and the $L$-test in almost every simulation setting of heterogeneity, aggregation and
regularity, especially when the sample size is large.
 \end{modnew}  \begin{modnew}
     We further verify the robustness of the CF test to boundary points in the Appendix.
 \end{modnew}

\section{Testing procedure}

\begin{modnew}
    \subsection{General form}
\end{modnew}

In the scope of this paper, we define a spatial point process as a set of $n$ random vectors (or points) $\{\bm{x}_1, \ldots, \bm{x}_n;\; \bm{x}_j \in \mathcal{A} \subset \mathbb{R}^D\}$, where $\mathcal{A}$ is  \begin{modnew}
    a
\end{modnew} study region of finite size. Most often, $\mathcal{A}$ is a rectangle and can be transformed linearly into the unit square $[0, 1]^2$. The sample size $n$ is considered fixed and the vectors are exchangeable in indices. There may be dependence between the points. Then, CSR is the special case where the points are mutually independent and all follow the uniform distribution in $\mathcal{A}$. 

To test a null assumption like CSR using the CF, a popular choice of statistic is a weighted $L_2$-like distance, scaled by $n$,
\begin{align*}
    \Delta &= \int_{\mathbb{R}^D} n\bigl|\phi_0(\bm{t}) - \hat{\phi}_X(\bm{t})\bigr|^2 w(\bm{t})\,d\bm{t},
\end{align*}
where $\hat{\phi}_X(\bm{t}) = (1/n)\sum_{j=1}^n \exp(i\bm{t}^\intercal \bm{x}_j)$ is the empirical CF, and $\phi_0(\bm{t})$ is the CF of the null distribution \citep{JIMENEZGAMERO20093957}, which is the uniform distribution in $\mathcal{A}$ in our case. If the marginal CF is truly $\phi_0$, then $\hat{\phi}_X(\bm{t})$ is an unbiased estimator of it, i.e., $E_0 \{\phi_0(\bm{t}) - \hat{\phi}_X(\bm{t})\} = 0$.

The weight function $w(\bm{t})$ is usually chosen as the squared modulus of the CF, $|\phi_0(\bm{t})|^2$, \begin{modnew}
    or the
\end{modnew} Gaussian  or double exponential (Laplace) density \citep{Meintanis}, but we  find it more convenient to use the Cauchy density when testing for CSR \begin{modnew}
    because it is then possible to derive the asymptotic null distribution of the test statistic
\end{modnew}. 

 \begin{modnew}
    A\ignorespacesafterend
\end{modnew}s shown by \citet{JIMENEZGAMERO20093957}, \begin{modnew}
    for an arbitary weight function $w(\bm{t})$,
\end{modnew}
\begin{align}
    \Delta &= \int_{\mathbb{R}^D} \frac{1}{n} \sum_{j, k = 1}^n \Bigl\{e^{i\bm{t}^\intercal\bm{x}_j}-E \bigl(e^{i\bm{t}^\intercal\bm{y}}\bigr)\Bigr\}\Bigl\{e^{-i\bm{t}^\intercal\bm{x}_k}-E \bigl(e^{-i\bm{t}^\intercal\bm{y}}\bigr)\Bigr\}w(\bm{t})\,d\bm{t} \nonumber\\
    &= \frac{1}{n} \sum_{j, k = 1}^n \xi(\bm{x}_j - \bm{x}_k) - 2\sum_{j = 1}^n E_{\bm{x}_j} \bigl\{\xi(\bm{x}_j - \bm{y})\bigr\} + nE \bigl\{\xi(\bm{y} - \bm{y}')\bigr\},
    \label{eq:test_stat}
\end{align}
where $\bm{y}$ and $\bm{y}'$ are placeholders that have $D$-dimensional uniform distributions and are independent of each other and $(\bm{x}_1, \ldots, \bm{x}_n)$, and the function 
\[ \xi(\bm{x}) = \int_{\mathbb{R}^D} e^{i \bm{t}^\intercal \bm{x}} w(\bm{t})\, d\bm{t} \]
is the  \begin{modnew}
    Fourier dual of
\end{modnew} the weight \begin{modnew}
    function
\end{modnew}  $w(\bm{t})$. \begin{modnew}
    If $w(\bm{t})$ is a probability density function, then $\xi(\bm{x})$ is its characteristic function.
\end{modnew} The subscript in $E_{\bm{x}_j}(\cdot)$ means that $\bm{x}_j$ is fixed.

The interpretation of Equation \eqref{eq:test_stat} is straightforward. The first term is the most important and concerns the distances between  every pair of points. 
\begin{modnew}
    It remains the same regardless of the null hypothesis and study region.
\end{modnew}
The second term captures the location of every point independently and  \begin{modnew}
    changes according to the null hypothesis and study region.
\end{modnew} The third term is a constant.
\begin{modnew}
    The first term is computationally the most expensive, with complexity $O(n^2)$ compared to $O(n)$ for the second term.
\end{modnew}
We can generate the test statistic forms we want with much flexibility by choosing $\xi$ and transforming it back to density $w$, using the inversion formula \citep{shephard1991a} or P\'olya’s theorem \citep{polya1949remarks, sato2004remarks}.

\begin{modnew}
    \subsection{Choice of the weight function} \label{chap:weight}
\end{modnew}

Later simulations will show that the effective range or scale of the weight function $w(\bm{t})$ has great influence on the power of the CF test for detecting  spatial interactions \begin{modnew}
    at different distances
\end{modnew}, so we convert the weight function to a parameterized form, \begin{modnew}
    $w_r(\bm{t}) = r^{D}w(r\bm{t})$ and corresponding $\xi_r(\bm{x}) = \xi(\bm{x}/r)$, with a positive scale parameter $r$. The parameter $r$ will also have an important role parallel to its role in Ripley's $K(r)$.
\end{modnew}  
\begin{modnew}
    Here, we list some reasonable choices of the weight function with the scale parameter $r$ incorporated.
    \begin{enumerate}
        \item Bessel-like weight function. In one dimension, there is a weight function that produces a simple rectangular $\xi_r$,
        \[ w_r(t) = \frac{\sin(rt)}{\pi t}, \quad \xi_r(x) = 1(|x| \le r).\]
        We can extend it to $D$ dimensions so that $\xi_r$ is an indicator function of a $D$-dimensional ball centered at the origin,
        \[ w_r(\bm{t}) = ||2\pi\bm{t}/r||^{-D/2} J_{D/2}(||r\bm{t}||), \quad \xi_r(\bm{x}) = 1(||\bm{x}|| \le r), \]
        where $J_{D/2}(\cdot)$ is the Bessel function of order $D/2$. This is conceptually very similar to Ripley's $K$-function. In fact, according to Equation \eqref{eq:null_expectation} derived later, the expectation of the CF test statistic under the null hypothesis is $E_0(\Delta_r) = 1 - E_0 \{\xi_r(\bm{x}_1 - \bm{x}_2)\} = 1 - \mathrm{pr}(||\bm{x}_1 - \bm{x}_2|| \le r)$.
        Therefore, $E_0(1 - \Delta_r)$ is the probability (or expected proportion) of other points falling inside a disk of radius $r$ centered at a randomly distributed point, and $|\mathcal{A}|(1 - \Delta_r)$ is an unbiased estimator of a boundary-corrected (or nonstationary) version of Ripley's $K$-function. (Note that this estimator is only unbiased conditional on the sample size, as are most estimators of Ripley's $K$-function.) For example, when the study region $\mathcal{A} = [0, 1]^2$, $E_0(1 - \Delta_r) = \pi r^2 - 8/3\cdot r^3 + 1/2 \cdot r^4$ for $r \in (0, 1]$ \citep{MathSE}, similar to the original Ripley's K function $K(r) = \pi r^2$ and asymptotically equivalent as $r \rightarrow 0$. The corresponding CF test statistic is
        \begin{align*}
            \Delta_r =& \frac{1}{n} \sum_{j, k = 1}^n 1\left(||\bm{x}_j - \bm{x}_k|| < r\right) - 2\sum_{j=1}^n \mathrm{pr}_{\bm{x}_j}(||\bm{x}_j - \bm{y}|| < r) + n\left\{1-\pi r^2 + \frac{8}{3} r^3 - \frac{1}{2} r^4\right\}.
        \end{align*}
        Its first term can be compared to the empirical Ripley's $K$-function \citep{baddeley2015spatial},
        \[ \hat{K}(r) = \frac{|\mathcal{A}|}{n(n-1)} \sum_{\substack{j, k=1\\ j\neq k}}^n  1\left(||\bm{x}_j - \bm{x}_k|| < r\right) e(\bm{x}_j,\bm{x}_k,r), \]
        where $e(\bm{x}_j,\bm{x}_k,r)$ can be one of several types of edge correction weight.
        The expression $\mathrm{pr}_{\bm{x}_j}(||\bm{x}_j - \bm{y}|| < r)$ in $\Delta_r$ is the proportion of the square $[0, 1]^2$ that is covered by a disk of radius $r$ centered at $\bm{x}_j$. For instance, if the point $\bm{x}_j$ is at least distance $r$ from the boundary, $\mathrm{pr}_{\bm{x}_j}(||\bm{x}_j - \bm{y}|| < r) = \pi r^2$.
        \item The normalized squared modulus of the null distribution CF $\phi_0(\bm{t})$ is a common choice for weight function in the CF-based test \citep{JIMENEZGAMERO20093957}. For the uniform null distribution in the CSR problem,
        \[ w(\bm{t}) = \frac{1}{(2\pi)^D} |\phi_0(\bm{t})|^2 = \prod_{d=1}^D\frac{1 - \cos(t_d)}{\pi t_d^2}, \quad \xi(\bm{x}) = \prod_{d=1}^D(1 - |x_d|)^+.\]
        The $\xi$ function has a simple triangular shape.
        When the test region is $[0,1]^2$, we actually recover \citet{zimmerman1993}'s $\bar{\omega}^2$ statistic for CSR, multiplied by a factor of $4$:
\begin{align*}
    \Delta &= \frac{1}{n}\sum_{j, k = 1}^n \bigl(1-|x_{j1} - x_{k1}|\bigr)\bigl(1-|x_{j2} - x_{k2}|\bigr) - 2\sum_{j = 1}^n \Bigl(x_{j1}^2 - x_{j1} - \frac{1}{2}\Bigr)\Bigl(x_{j2}^2 - x_{j2} - \frac{1}{2}\Bigr) + \frac{4}{9}n\\
    & = 4\bar{\omega}^2.
\end{align*}
        \item Centered Gaussian weight function,
        \[ w_r(\bm{t}) = \left(\frac{r}{\sqrt{\pi}}\right)^D \exp\left(-r^2\sum_{d=1}^Dt_d^2\right), \quad \xi_r(\bm{x}) = \exp\left(-\frac{1}{r^2} \sum_{d=1}^D x_d^2\right) .\]
        The corresponding statistic in $[0, 1]^D$ is
        \begin{align*}
    \Delta_r &= \frac{1}{n}\sum_{j, k = 1}^n \exp \left[-\frac{1}{r^2} \sum_{d=1}^D \left(x_{jd}-x_{kd}\right)^2\right] - 2  \sum_{j=1}^n \prod_{d=1}^D \frac{\sqrt{\pi}}{2} r \left[\text{erf}\Bigl(\frac{x_{jd}}{r}\Bigr)+\text{erf}\Bigl(\frac{1-x_{jd}}{r}\Bigr)\right] \\
    &\quad +  n\left[r\left\{\sqrt{\pi}\text{erf}\Bigl(\frac{1}{r}\Bigr) + r e^{-1/r^2} - r\right\} \right]^D.
\end{align*}
        Being the only weight function that is both isotropic and separable, the Gaussian weight function may be valuable in high dimensions, where other separable weight functions can be grossly anisotropic.
        \item Cauchy weight function,
        \begin{equation}
        \label{eq:cauchy_weight}
            w_r(\bm{t}) = \prod_{d=1}^D\frac{r}{\pi \bigl\{1+(rt_d)^2\bigr\}}, \quad \xi_r(\bm{x}) =  \exp\left(-\frac{1}{r}\sum_{d=1}^D|x_d|\right).
        \end{equation}
        The corresponding statistic in $[0, 1]^D$ is
        \begin{align*}
    \Delta_r &= \frac{1}{n}\sum_{j, k =1}^n \exp\left(-\frac{1}{r}\sum_{d=1}^D|x_{jd} - x_{kd}|\right) - 2 \sum_{j=1}^n\prod_{d=1}^D r\Bigl\{2 - e^{-x_{jd}/r} - e^{-(1-x_{jd})/r}\Bigr\}\\
    &\quad+ n\left\{2r\left(1+re^{-1/r}-r\right)\right\}^D.
\end{align*}
    \end{enumerate}
\end{modnew}

\begin{modnew}
While tests based on each one of the above statistics can be implemented via Monte Carlo simulation, we are able to develop efficient algorithms for the asymptotic and small-sample null distributions of the Cauchy-weighted statistic in Section \ref{chap:null_distribution}. The calculation of this statistic is also very simple. Therefore, it will be our main focus from now on. 
\end{modnew}

\subsection{Details of the Cauchy-weighted CF test} \label{chap:cauchy}

With a Cauchy weight function, the quantity $\xi(\bm{x}_j - \bm{x}_k)$ in Equation \eqref{eq:test_stat} of the test statistic is simply the exponential of  the $L^1$ distance between $\bm{x}_j$ and $\bm{x}_k$ \begin{modnew}
    divided by $r$
\end{modnew}.
The  \begin{modnew}
    smaller the parameter $r$
\end{modnew}, the more peaked the function $\xi$. As a result, the test statistic will become more sensitive to (by having more variation in) smaller distances of spatial interaction while losing strength at larger distances. Therefore,  \begin{modnew}
    the parameter $r$ roughly represents the distance of spatial interaction, as we have already seen in the Bessel-weighted CF test
\end{modnew}. \begin{modnew}
    We can thus predict that heterogeneity, a large-scale phenomenon, is better detected by larger values of $r$, while local features like aggregation and regularity are more likely to show up at smaller $r$.
\end{modnew}

 \begin{modnew}
    The distances between neighboring points should scale with $n^{-1/D}$ under the null distribution. Because of this and some empirical testing for $D = 2, 3$, we recommend choosing $r \in [n^{-1/D}/(4\pi), 1]$ for the CF test in the study region $[0, 1]^D$. If there are no preferable values for $r$, a graph plotting the test statistic against $r$ for $r\in [n^{-1/D}/(4\pi), 1]$ 
\end{modnew} is a useful visual substitute for the test statistic\begin{modnew}
    , similar to the envelope graph of Ripley's $K$-function
\end{modnew}. We will demonstrate its usage in Section \ref{chap:application}. 

Alternatively, we can use an omnibus test that combines the test statistics corresponding to several  \begin{modnew}
    $r$
\end{modnew} values. Simply using the Bonferroni correction will suffice: given the $p$-values $p_1$, $p_2$, $\ldots$, $p_m$ from multiple CF tests, the Bonferroni-corrected overall $p$-value is 
\[ \min\Bigl(m \min_{i=1}^m p_i, \; 1\Bigr). \]
The validity of the omnibus CF test is guaranteed no matter the dependence structure between the $p$-values. Nonetheless, the values of  \begin{modnew}
    $r$
\end{modnew} should be sparsely spaced so that there is less correlation and hence more power. One possibility is to choose  \begin{modnew}
    $r$ to be the three values $(4\pi n^{1/D})^{-1}$, $(4\pi n^{1/D})^{-1/2}$, 1 for $D = 2, 3$
\end{modnew}, which will be shown to have good performance in the simulation\begin{modnew}
    s
\end{modnew} and application.

\begin{modnew}
    The computational complexity of the Cauchy-weighted statistic is $O(n^2)$, so if the number of points is very large, the calculation may be computationally prohibitive. However, after optimizing our implementation (in Rust code) of the Cauchy-weighted test statistic with parallelism and `single instruction, multiple data' operations, this does not appear to be a problem in the vast majority of cases. For example, when tested on a personal computer (a common 64-bit Windows machine with a 6-core CPU), it takes about $0.5$ second to compute the statistic for a two-dimensional CSR pattern of $50,000$ points, and 10 seconds for a CSR pattern of $200,000$ points. The computation time of three-dimensional patterns is only slightly more. 
    
    Even when the sample size is so large that the computation of the Cauchy-weighted CF statistic is significantly slowed down, this is less of a problem than it is for other CF tests and the $L$-test, since we can obtain the null distribution theoretically for the Cauchy-weighted statistic with little time cost instead of using a slow Monte Carlo simulation. For example, the theoretical $p$-value calculation for the Bonferroni combination of three CF tests as stated earlier takes about $1.8$ second for a CSR sample of size $50,000$, and $30$ seconds for a CSR sample of size $200,000$, while the simulation based $p$-value for the isotropic-corrected $L$-test, obtained from only $99$ Monte Carlo simulations (and thus quite inaccurate), takes about $108$ and $546$ seconds respectively. 
\end{modnew}

\section{CF test under the alternative hypotheses}

The CF test based on a weighted $L_2$-like distance is conventionally used for detecting heterogeneity. According to Theorem 2 in \citet{JIMENEZGAMERO20093957}, if the points $\bm{x}_j$ are mutually independent, the test is strongly consistent against any fixed alternative (non-uniform) distribution, i.e., $\Delta \rightarrow \infty$ a.s. as sample size $n \rightarrow \infty$. It is also able to detect alternatives that differ from the null by $O(n^{-1/2})$, in the sense that the asymptotic distribution of $\Delta$ is different from that of the uniform case.

Although theoretically valid, no investigation has been made on the quality of this type of CF test against aggregation and regularity. We will illustrate how such alternatives would bias the expectation of the test statistic,
\begin{align*}
    E (\Delta) = \frac{1}{n} \sum_{j, k = 1}^n E\bigl\{\xi(\bm{x}_j - \bm{x}_k)\bigr\} - 2\sum_{j = 1}^nE \bigl\{\xi(\bm{x}_j - \bm{y})\bigr\} + nE \bigl\{\xi(\bm{y} - \bm{y}')\bigr\}.
\end{align*}

Write $\elemabs{\bm{x}}$ as the element-wise absolute value of $\bm{x}$. Expression $\xi(\bm{x})$ is a function of $\elemabs{\bm{x}}$ provided that the density $w(\bm{t})$ is symmetric. We additionally require $\xi(\bm{x})$ to be nonincreasing with respect to $\elemabs{\bm{x}}$, i.e., $\xi(\bm{x})\ge\xi(\bm{y})$ if $\bm{0}\le\bm{x}\le\bm{y}$, which holds for the Cauchy CF (here the order $\bm{x} \le \bm{y}$ holds if and only if all of their components $x_d \le y_d$). 

To simplify our analysis, we consider aggregation and regularity to be results of dependence structure between the points, and individually the points still follow the uniform distribution in the study region $\mathcal{A}$. Additionally, we introduce the notion of stochastic order. Given two random vectors $\bm{X}$ and $\bm{Y}$, $\bm{X}$ is said to be stochastically smaller than $\bm{Y}$ if their tail probabilities satisfy $\mathrm{pr}(\bm{X}>\bm{x}) \le \mathrm{pr}(\bm{Y}>\bm{x})$ for any $\bm{x}$ \citep{shaked2007stochastic}. Roughly speaking, this means that $\bm{X}$ is less likely to take on large values.

We define aggregation as the situation where every pairwise difference in absolute values, $\elemabs{\bm{x}_j-\bm{x}_k} \ (j \neq k)$, is stochastically strictly smaller than that of independent uniform variables, $\elemabs{\bm{y}-\bm{y}'}$. This is a very reasonable characterization and at least captures various kinds of Poisson cluster process. \citet{shaked2007stochastic} showed that the stochastic ordering implies that $E \{\xi(\bm{x}_j-\bm{x}_k)\} \ge E \{\xi(\bm{y}-\bm{y}')\}$ for $\xi(\bm{x})$ nonincreasing  \begin{modnew}
    with respect to
\end{modnew} $\elemabs{\bm{x}}$. Therefore, compared to the CSR case, aggregation increases the $E \{\xi(\bm{x}_j - \bm{x}_k)\}$ term and thus the expectation of $\Delta$. The other two terms are unchanged as individual $\bm{x}_j$ still follow a uniform distribution.

Conversely, a point process is regular when every pairwise difference in absolute values is stochastically strictly greater than that of independent uniform variables. This results in a smaller $E \{\xi(\bm{x}_j - \bm{x}_k)\}$ term and smaller expectation of $\Delta$.

\section{\begin{modnew}
    Distribution of the
\end{modnew} CF test \begin{modnew}
    statistic
\end{modnew} } \label{chap:null_distribution}

\subsection{Mean and variance} \label{chap:first_two_moments}

Under the null hypothesis that the samples $\{\bm{x}_1, \ldots, \bm{x}_n\}$ are independent and all follow the uniform distribution in $\mathcal{A}$, it is relatively easy to obtain the mean and variance of the test statistic.
\begin{proposition} \label{prop_first_two_moments}
    If the CF associated with the chosen weight function $w(\bm{t})$ is $\xi(\bm{x})$, the mean and variance of the test statistic $\Delta$ under the null hypothesis are
    \begin{align}
        E_0(\Delta) &= 1 - E_0 \{\xi(\bm{x}_1 - \bm{x}_2)\}, \label{eq:null_expectation}\\
        \mathrm{var}_0 (\Delta) &= \frac{2n-6}{n}[E_0\{\xi(\bm{x}_1 - \bm{x}_2)\}]^2 + \frac{2n-2}{n} E_0\left[\{\xi(\bm{x}_1 - \bm{x}_2)\}^2\right]\nonumber\\
            &\quad\, - \frac{4n-8}{n} E_0 \{\xi(\bm{x}_1 - \bm{x}_3)\xi(\bm{x}_2 - \bm{x}_3)\}. \label{eq:null_variance}
    \end{align}
\end{proposition}
A proof is given in the Appendix.

When $w(\bm{t})$ is the Cauchy weight density in Equation \eqref{eq:cauchy_weight}, the mean and variance are
\begin{modnew}
    \begin{align*}
    E_0(\Delta) &= 1 - \left\{2r(1+re^{-1/r}-r)\right\}^D,\\
    \mathrm{var}_0 (\Delta) &= \frac{2n-6}{n}\left\{2r(1+re^{-1/r}-r)\right\}^{2D} + \frac{2n-2}{n} \left\{r\left(1+\frac{1}{2}re^{-2/r}-\frac{1}{2}r\right)\right\}^D\\
    &\quad\, - \frac{4n-8}{n} \left\{r^2\left(4 + 2e^{-1/r} + 8re^{-1/r} - re^{-2/r} - 7r\right)\right\}^D.
\end{align*}
\end{modnew}

In fact, moments of any order may be obtained in closed form in this case. This is not true at least for the variance when other weights like Gaussian or double exponential density functions are used.

\subsection{Asymptotic distribution \begin{modnew}
    under Cauchy weights in a square study region
\end{modnew}}

As shown in \citet{JIMENEZGAMERO20093957}, theorem 6.4.1.B in \citet{serfling1980approximation} can be used to derive the asymptotic distribution of $\Delta$. Specifically, $\Delta$ converges in distribution to $\sum_{j=1}^\infty \lambda_j Z^2_{j}$, where $(Z_{j})$ is a sequence of  independent standard normal variables and $(\lambda_j)$ are the eigenvalues of the linear operator $T: \mathcal{L}^2(\mathcal{A}) \rightarrow \mathcal{L}^2(\mathcal{A})$ of square-integrable functions defined by
\begin{equation}
    \label{eq:linear_operator}
    Tg(\bm{x}) = \frac{1}{\int_{\mathcal{A}} d\bm{y}}\int_{\mathcal{A}} [\xi(\bm{x} - \bm{y}) - E \{\xi(\bm{x} - \bm{y}')\} - E \{\xi(\bm{x}' - \bm{y})\} + E \{\xi(\bm{x}' - \bm{y}')\}] g(\bm{y}) \,d \bm{y}.
\end{equation}
Here, $\bm{x}'$ and $\bm{y}'$ are uniform random  \begin{modnew}
    vectors
\end{modnew} in $\mathcal{A}$.

Finding the eigenvalues of this operator is in general a very difficult task

\begin{modnew}
    because it is often not possible to find an appropriate set of basis functions that can give a tidy matrix form for an operator. Fortunately, the eigenvalues can be obtained using basic linear algebra
\end{modnew} when  independent Cauchy weights \begin{modnew}
     are used
 \end{modnew} and the study region $\mathcal{A} = [0, 1]^D$.
 \begin{modnew}
     This is because the operator then has a special matrix form that is several rank-one and -two updates away from a diagonal matrix. In the Appendix, we develop an efficient algorithm to find the asymptotic distribution of the test statistic.
 \end{modnew}

Simulations show that when the  \begin{modnew}
    parameter $r$ is small
\end{modnew}, the distribution of the test statistic

converges too slowly to the asymptotic distribution \begin{modnew}
    that we have developed
\end{modnew}. Adjusting the obtained quantile $q$ using the exact variance, i.e.,
\[ \tilde{q} = \{q - E (\Delta)\} \left\{\frac{\mathrm{var}(\Delta)}{\lim_{n \rightarrow \infty} \mathrm{var} (\Delta)}\right\}^{1/2} + E (\Delta), \]
will offer a slight improvement, but not a complete fix. In the next section, we  demonstrate an alternative distribution for small sample sizes.

\subsection{Small-sample distribution at  \begin{modnew}
    small $r$
\end{modnew}}

\begin{modnew}
     To fix the slow convergence rate of the asymptotic distribution, w
\end{modnew}e \begin{modnew}
    also \
\end{modnew} propose an alternative method that gives a small\begin{modnew}
    -
\end{modnew}sample distribution that is asymptotically exact as  \begin{modnew}
    $r \rightarrow 0$
\end{modnew}.

Since the sample size $n$ is now assumed to be finite, the test statistic has bounded support. Therefore, according to the Hausdorff moment problem, the moments should completely determine the distribution. To find the distribution using the moments, we can first find its CF. The logarithm of the CF, the cumulant generating function\begin{modnew}
    ,
\end{modnew} is
\[  \ \begin{modnew}
    \psi
\end{modnew}(t) = \log\left\{E (e^{it\Delta})\right\} = \sum_{m=1}^\infty \frac{\kappa_m}{m!}(it)^m,  \]
where $\kappa_m =  \ \begin{modnew}
    \psi
\end{modnew}^{(m)}(0)/i^m$ is the $m$th cumulant. We \begin{modnew}
    now
\end{modnew} give a simple formula for the cumulants when  \begin{modnew}
    $r \rightarrow 0$
\end{modnew}.

\begin{proposition}
    \label{asymptotic_cumulant}
     \begin{modnew}
    As $r \rightarrow 0$
\end{modnew}, the $m$th cumulant of test statistic $\Delta$ satisfies
    \[ \kappa_m = (n-1)\left(\frac{2}{n}\right)^{m-1}\left(\frac{2}{m}\right)^D\begin{modnew}
        r^D
    \end{modnew} \begin{modnew}
        + o(r^D)
    \end{modnew}  \quad (m \ge 2). \]
\end{proposition}
A proof is given in \begin{modnew}
    the
\end{modnew} Appendix.

Therefore, the cumulant generating function of $\Delta$ has expansion $ \ \begin{modnew}
    \psi
\end{modnew}(t) =  \ \begin{modnew}
    \psi
\end{modnew}_0(t)\begin{modnew}
        + o(r^D)
    \end{modnew} $, where
\[  \ \begin{modnew}
    \psi
\end{modnew}_0(t) = it\kappa_1 + \sum_{m=2}^\infty (n-1)\left(\frac{2}{n}\right)^{m-1}\left(\frac{2}{m}\right)^D\begin{modnew}
        r^D
    \end{modnew}\frac{(it)^m}{m!}, \quad \kappa_1 = E(\Delta). \]
It is easy to verify that the series expansion of $ \ \begin{modnew}
    \psi
\end{modnew}_0(t)$ converges for every $t \in \mathbb{R}$, but there is no guarantee that $ \ \begin{modnew}
    \psi
\end{modnew}_0(t)$ corresponds to an actual random variable (which requires that $\exp( \ \begin{modnew}
    \psi
\end{modnew}_0(t))$ is a positive definite function of $t$). For the  \begin{modnew}
    values of $r$
\end{modnew} and $n$ we tested, it does appear to be the case. 

This allows us to find an asymptotic CDF of the test statistic  \begin{modnew}
    as $r \rightarrow 0$
\end{modnew}, via the inversion formula of \citet{gil-pelaez},
\begin{align*}
    \mathrm{pr} (\Delta\le x) &= \frac{1}{2}-\frac{1}{\pi}\int_0^\infty \frac{\im\bigl[\exp\{ \ \begin{modnew}
    \psi
\end{modnew}(t)-itx\}\bigr]}{t}\,dt\\
    &= \frac{1}{2}-\frac{1}{\pi}\int_0^\infty \frac{\sin\bigl[\im\{ \ \begin{modnew}
    \psi
\end{modnew}_0(t)\}-tx\bigr]}{t} \exp\bigl[\re\{ \ \begin{modnew}
    \psi
\end{modnew}_0(t)\}\bigr] \,dt \begin{modnew}
        + o(r^D)
    \end{modnew} ,
\end{align*}
where functions $\re(\cdot)$ and $\im(\cdot)$ denote the real and imaginary parts of a complex number, respectively. Technically, the last equality requires that the remainder \begin{modnew}
        $o(r^D)$
    \end{modnew}  in the expansion of $ \ \begin{modnew}
    \psi
\end{modnew}(t)$ does not grow too fast with $t$, so we also verify this equation through simulation in Section \ref{chap:sim_null_distribution}. The integral in the equation can be evaluated numerically using the Gauss-Konrod quadrature algorithm from the C++ Boost package. We can also calculate the distribution quantiles using the `bracket and solve' root-finding method in the same package and adjust the results using the exact variance as in the last section.

Whether to use the asymptotic distribution corresponding to $n \rightarrow \infty$ or  \begin{modnew}
    $r \rightarrow 0$
\end{modnew} should depend on the  \begin{modnew}
    size of the product $n^{1/D} r$
\end{modnew}. The simulation in Section \ref{chap:sim_null_distribution}  \begin{modnew}
    for which
\end{modnew} dimension $D=2$ shows that when  \begin{modnew}
    $n^{1/2}r \approx 1/\pi $ or equivalently $r \approx (\pi n^{1/2})^{-1}$
\end{modnew}, the two methods give approximately the same distribution. When  \begin{modnew}
    $r < (\pi n^{1/2})^{-1}$
\end{modnew}, the small\begin{modnew}
    -
\end{modnew}sample distribution for  \begin{modnew}
    $r \rightarrow 0$
\end{modnew} should be used. Otherwise, the large\begin{modnew}
    -
\end{modnew}sample asymptotic distribution for $n \rightarrow \infty$ is better. This has not been tested  \begin{modnew}
    for
\end{modnew} other $D$.

\section{Simulation} \label{chap:simulation}

\subsection{Accuracy of the proposed null distribution} \label{chap:sim_null_distribution}

\begin{figure}
    \figuresize{.7}
    \figurebox{20pc}{25pc}{}[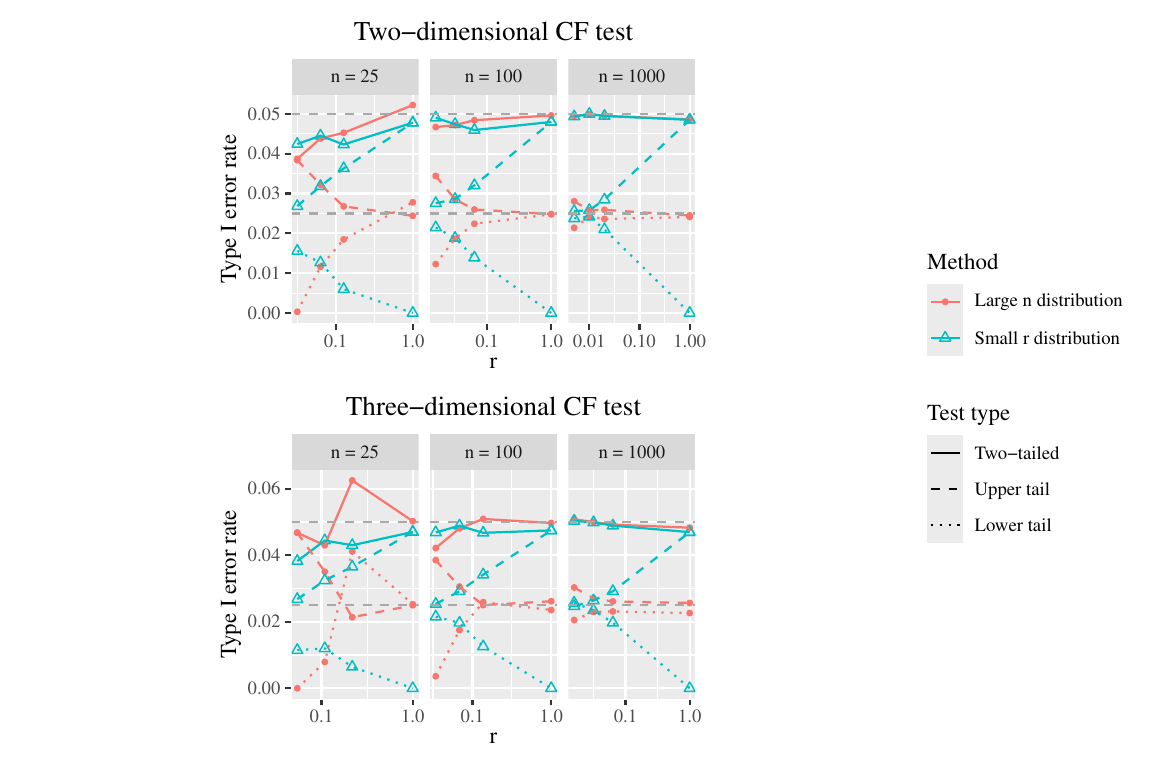]
    \caption{The estimated type I error rate \begin{modnew}
        of both the two- and three-dimensional Cauchy-weighted CF tests
    \end{modnew} when using the two proposed null distributions. The theoretical test size is 0.025 for the one-tailed tests, and 0.05 for the two-tailed test.}
    \label{fig:type1_error}
\end{figure}

To assess the accuracy of the two proposed null distributions for the test statistic, we simulate data from the uniform distribution in \begin{modnew}
    both
\end{modnew} the unit square $[0, 1]^2$ \begin{modnew}
    and the unit cube $[0, 1]^3$,
\end{modnew} and estimate the type I error rate. The simulation repeats $5 \times 10^4$ times. The test size is 0.05 and sample size $n$ is  \begin{modnew}
    25, 100, 1000
\end{modnew}. To ensure that the asymptotic null distribution is accurate in each tail, one-tailed tests with half the test size are also included. The  \begin{modnew}
    parameter $r \in \{1/2 \cdot(\pi n^{1/D})^{-1}, (\pi n^{1/D})^{-1}, 2 \cdot (\pi n^{1/D})^{-1}, 1\}$ is a function of sample size $n$, where $D = 2, 3$ is the dimension
\end{modnew}. The results are given in Fig. \ref{fig:type1_error}.

As we can see, the two-tailed test using either null distribution method has type I error rate very close to the theoretical value 0.05, regardless of  \begin{modnew}
    $r$
\end{modnew}, but this error rate can be very unbalanced between the upper and lower tail. It appears that for a fixed $n$, as  \begin{modnew}
    $r$ decreases toward zero
\end{modnew}, the  \begin{modnew}
    small $r$
\end{modnew} distribution becomes more accurate, while the large $n$ distribution becomes less \begin{modnew}
    so
\end{modnew}. When  \begin{modnew}
    $r \approx (\pi n^{1/D})^{-1}$
\end{modnew}, they have  similar performance. The type I error rate also seems to become more \begin{modnew}
    accurate and
\end{modnew} balanced in the two tails as the sample size $n$ increases , \begin{modnew}
    for every value of $r$,
\end{modnew} which is good, as \begin{modnew}
    the computational cost of
\end{modnew} simulating the null distribution is  \begin{modnew}
    larger
\end{modnew} for large $n$.

\subsection{ \begin{modnew}
    Power
\end{modnew} of the CF test} \label{chap:power}

Next, we perform simulations on the CF test\begin{modnew}
    's
\end{modnew} power under different alternatives to CSR, compared with the $L$-test, the Clark-Evans test and Zimmerman's $\bar{\omega}^2$-test. The CF test uses Cauchy density weights with   \begin{modnew}
    $r = (4\pi n^{1/2})^{-1}, (4\pi n^{1/2})^{-1/2}, 1$, as we have recommended in Section \ref{chap:cauchy}
\end{modnew}. We also include the Bonferroni-corrected omnibus CF test \begin{modnew}
    that combines the three individual CF tests with different values of $r$
\end{modnew}. The $L$-test statistic is calculated via $L_m = \sup_{r \le s} |\{\hat{K}(r) / \pi\}^{1/2} - r|$, where $\hat{K}(r)$ is \citet{Ripley1988}'s isotropic\begin{modnew}
    -edge
\end{modnew}-corrected estimator and $s = 1.25/n^{1/2}$ (in accordance with \citet{ripley1979}'s recommendation that $s$ should be $0.25$ for $n = 25$ and $0.125$ for $n = 100$). The Clark-Evans test statistic is evaluated with \citet{donnelly1978}'s edge correction.

The null hypothesis is taken to be two-dimensional CSR inside the unit square $[0, 1]^2$. The number of points $n$ for each point pattern is fixed to be  \begin{modnew}
    25, 100 or 1000
\end{modnew}. The size of the tests is $\alpha = 0.05$ and the corresponding (two-sided) thresholds under the fixed $n$'s are estimated beforehand via Monte Carlo simulations of  \begin{modnew} \ $10,000$ \ \end{modnew} CSR patterns.

The following three alternatives tested are similar to those tested by \citet{zimmerman1993}.
\begin{enumerate}
    \item \textbf{Matérn cluster process} \citep{Matérn1986}. This assesses the power of the tests under clustering. The Matérn cluster process first generates cluster centers as points of a Poisson process, and then each cluster center is replaced with a cluster of uniformly distributed points in a disk centered at it. The number of points in each cluster follows the Poisson distribution. Additionally, the simulation is conditioned on a fixed sample size  \begin{modnew}
        $n = 25, 100, 1000$
    \end{modnew} using rejection sampling \citep{Møller} (so that we only need to simulate the thresholds for the fixed $n$'s). The cluster radius is $ \begin{modnew}
        \delta
    \end{modnew} = 0.075, 0.15, 0.30$, and to add more difficulty, we reduce the number of points per cluster as cluster radius decreases: The mean number of points per cluster $\mu = n^{1/4}, n^{1/3}, n^{1/2}$ and the mean number of clusters $\kappa = n^{3/4}, n^{2/3}, n^{1/2}$ for $ \begin{modnew}
        \delta
    \end{modnew} = 0.075, 0.15, 0.30$ respectively.
    
    \setcounter{enumi}{1}
    \begin{modnew}
        \item \textbf{Diggle-Gates-Stibbard process} \citep{diggle1987, baddeley2015spatial}. This generating model simulates regularity and is a type of soft-core Gibbs process with pairwise potential function $c(r) = \sin^2\{(\pi/2)r/\rho\}$ for $r < \rho$ and $c(r) = 1$ for $r \ge \rho$, where $\rho$ is the maximum inhibition distance. In this process, a pairwise distance greater than $\rho$ is not penalized, but a distance below $\rho$ is penalized gradually as it becomes smaller.
    \end{modnew}
    \item \textbf{Inhomogeneous Poisson process}. For heterogeneity, an inhomogeneous Poisson process is simulated with intensity function $\lambda(x_1, x_2) = \{\theta_1 - (\theta_1-1)x_1\}\{\theta_2 - (\theta_2-1)x_2\}$.
\end{enumerate}

\begin{figure}
    \figuresize{.75}
    \figurebox{20pc}{25pc}{}[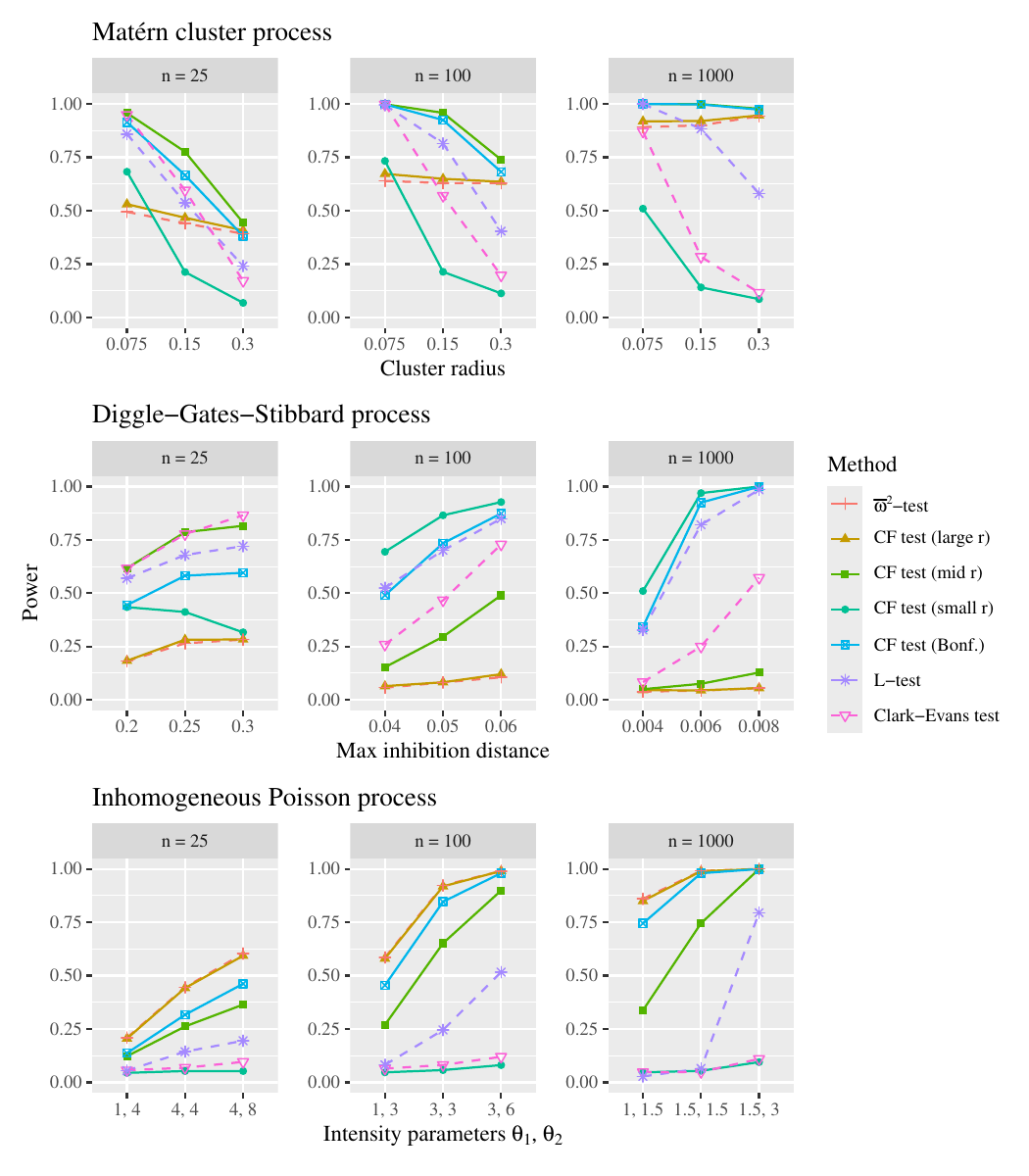]
    \caption{Simulation results of test power under different alternative hypotheses, comparing the CF test with $\bar{\omega}^2$-test, $L$-test and the Clark-Evans test. Non-CF tests are presented with dashed lines.}
    \label{fig:power}
\end{figure}

2000 independent realizations are simulated for each alternative \begin{modnew}
    using the \texttt{spatstat} package in \texttt{R}
\end{modnew}, to which the tests are applied and  power is estimated. The results are shown in Fig. \ref{fig:power}. The estimated power of the CF test with  \begin{modnew}
    $r = 1$
\end{modnew} is almost the same as \begin{modnew}
    that of
\end{modnew} Zimmerman's $\bar{\omega}^2$ in all cases. This is expected as they are both special cases of the CF test with very similar weight functions.

Individually, the CF test has greater power against spatial interactions at small distances when  \begin{modnew}
    the parameter $r$ is smaller
\end{modnew}, while losing power at  \begin{modnew}
    larger $r$
\end{modnew}. The omnibus CF test is always the second best among the CF tests (excluding Zimmerman's $\bar{\omega}^2$ since it is very similar to the CF test with $ \begin{modnew}
    r
\end{modnew} = 1$). It beats the \begin{modnew}
    $L$-test and
\end{modnew} Clark-Evans test in almost every scenario, the only exception being the  \begin{modnew}
    small-sample Diggle-Gates-Stibbard model
\end{modnew}. 
\begin{modnew}
    The omnibus CF test also seems to have a greater advantage as sample size increases.
\end{modnew}

\begin{modnew}
    In the Appendix, under similar simulation settings, we demonstrate that the omnibus CF has even greater power in the three-dimensional scenarios compared to the $L$-test.
\end{modnew}

\section{Application} \label{chap:application}

\begin{figure}
    \figuresize{.7}
    \figurebox{20pc}{25pc}{}[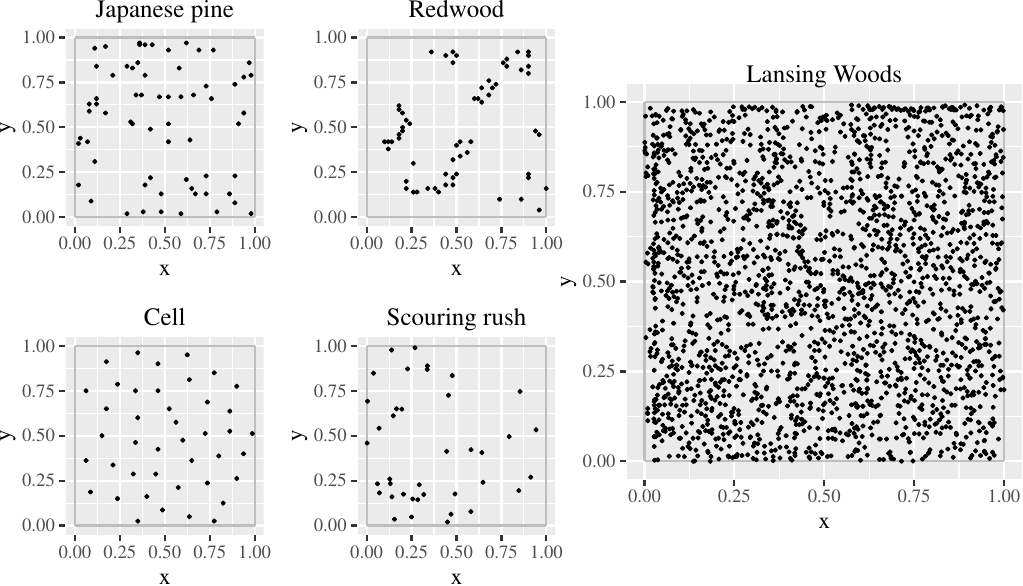]
    \caption{The  \begin{modnew}
        five
    \end{modnew} example spatial point patterns. Each dot represents a point in the pattern.}
    \label{fig:datasets}
\end{figure}

We apply the CF test to the four example point patterns used in \citet{zimmerman1993}:
\begin{enumerate}
    \item 65 Japanese black pine saplings in a square of side $5.7$ m \citep{numata1961};
    \item 62 redwood seedlings in a square of side $23$ m \citep{strauss1975};
    \item the centers of 42 biological cell in a unit square \citep{Ripley1977};
    \item 39 scouring rushes (Equisetum arvense) in a square of side 1.0 m, which were observed near Iowa City, Iowa \citep{zimmerman1993}.
\end{enumerate} \begin{modnew}
     These four patterns are supposed to demonstrate CSR, aggregation, regularity and heterogeneity, in that order. In addition, a point pattern with a larger sample size from the Lansing Woods dataset by \citet{gerrard1969competition} is tested. It records the location of 2250 trees (after removing one duplicated tree and not differentiated by species) in a 924 ft x 924 ft plot in Lansing Woods, Clinton County, Michigan USA.
\end{modnew}
The  \begin{modnew}
    five
\end{modnew} point patterns are shown in Fig. \ref{fig:datasets}, with their study regions standardized to the unit square $[0, 1]^2$. 

\begin{figure}
    \figuresize{.65}
    \figurebox{20pc}{25pc}{}[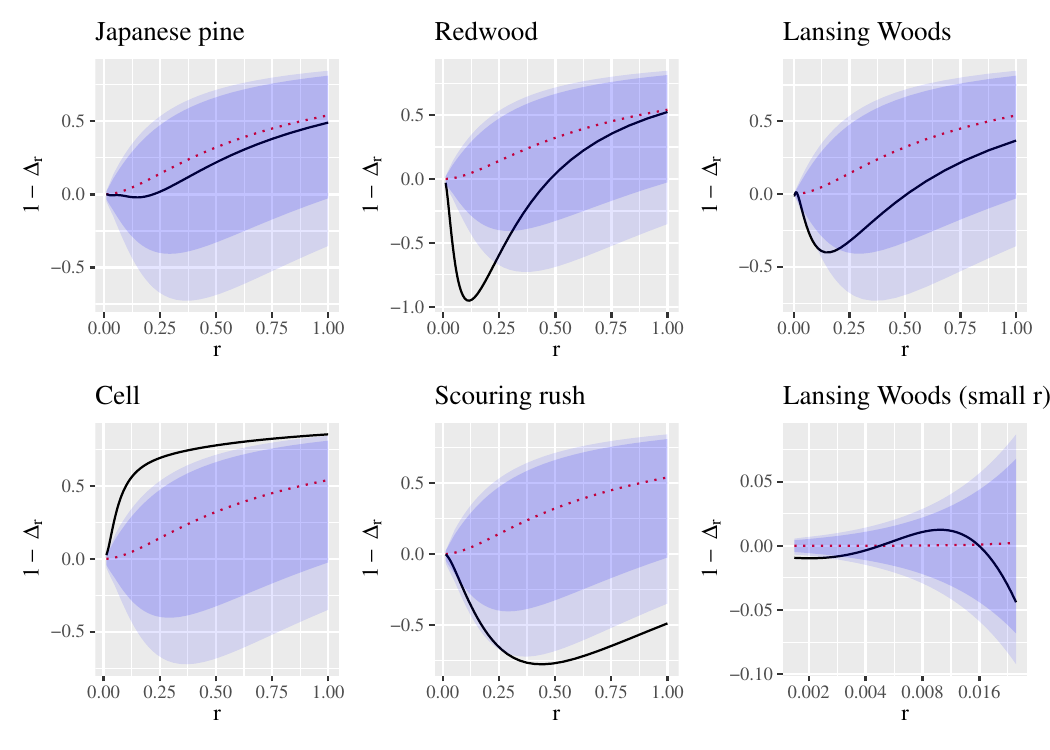]
    \caption{ \begin{modnew}
        The value of $1-\Delta_r$ in
    \end{modnew} the five example point patterns versus  \begin{modnew}
        the parameter $r$
    \end{modnew}, shown as the black line. \begin{modnew}
        An additional graph shows the statistics for the Lansing Woods point pattern at small $r$.
    \end{modnew} The red dotted line is the expected value of test statistic under null hypothesis. The dark and light blue areas contain the middle 95\% and 99\% of the null distribution, computed using the methods in Section \ref{chap:null_distribution}.}
    \label{fig:envelope}
\end{figure}

Figure \ref{fig:envelope} plots  \begin{modnew}
    $(1 - \Delta_r)$
\end{modnew} against the  \begin{modnew}
    parameter $r$
\end{modnew} using the null distribution developed in Section \ref{chap:null_distribution}. It resembles the envelope graph associated with the $L$-statistic, but without the need for simulations. The behavior of the test statistic is as expected: \begin{modnew}
    t
\end{modnew}he test statistics for the Japanese pine data are well within the middle 95\% of the null distribution\begin{modnew}
    ; t
\end{modnew}he heterogeneity in the scouring rush data induces a significant  \begin{modnew}
    negative
\end{modnew} departure \begin{modnew}
    (in $(1 - \Delta_r)$)
\end{modnew} from the null distribution at  \begin{modnew}
    large distance $r$
\end{modnew}\begin{modnew}
    ; and
\end{modnew} the aggregation in the redwood data does the same thing but at  \begin{modnew}
    smaller distance $r$
\end{modnew}. The regularity in the cell data leads to a general positive departure from the null distribution. For the Lansing Woods data, it is hard to spot any irregularities through human eyes, but its graph shows significant aggregation at two separate scales, one around $r = 0.1$, and another around $r = 0.002$. The latter is more significant.

We also apply the \begin{modnew}
    individual
\end{modnew} CF test\begin{modnew}
    s
\end{modnew}  and the Bonferroni-corrected omnibus CF test to the five example point patterns, compared with the $L$-test and the Clark-Evans test. The obtained significance levels, computed from Monte Carlo simulation  \begin{modnew}
    of $20,000$ patterns for the first four datasets and $2000$ patterns for the last one,
\end{modnew} are displayed in Table \ref{tab:p_value}.

\begin{table}
\def~{\hphantom{\textless}}
    \tbl{The significance level of the CF tests, $L$-test and Clark-Evans test applied to the five point patterns.  \begin{modnew}
        The standard error is less than $0.0036$ for the numbers in the first four data columns, and is less than $0.005$ for numbers less than $0.05$ in the last column
    \end{modnew}.}
{   \begin{tabular}{c | c c c c c}
        & \multicolumn{5}{c}{Point pattern}\\
        Method & Japanese pine & Redwood & Cell & Scouring rush & Lansing Woods\\
        \hline
        CF test, $r = 1$ & 0.627 & ~0.726 & ~0.005 & 0.004 & ~0.35 \\
        CF test, $r = (4\pi n^{1/2})^{-1/2}$ & 0.653 & \textless 0.001 & \textless 0.001 & 0.030 & ~0.02\\
        CF test, $r = (4\pi n^{1/2})^{-1}$ & 0.919 & ~0.076 & \textless 0.001 & 0.716 & \textless 0.01\\
        omnibus CF test & 1.000 & \textless 0.001 & \textless 0.001 & 0.013 & \textless 0.01\\
        $L$-test & 0.697 & \textless 0.001 & \textless 0.001 & 0.220 & \textless 0.01\\
        Clark-Evans test & 0.915 & \textless 0.001 & \textless 0.001 & 0.937 & ~0.04\\
    \end{tabular}}
    \label{tab:p_value}
\end{table}

Some of the individual CF tests are unable to detect departures in one of the scenarios, but the omnibus CF test successfully rejects CSR for the last four point patterns as expected. The $L$-test and the Clark-Evans test both fail to reject CSR for the scouring rush data set, even  \begin{modnew}
    though
\end{modnew} obvious heterogeneity is presented in the corresponding panel in Fig. \ref{fig:datasets}. \begin{modnew}
    For the large-sample Lansing Woods data set, the $L$-test and the small-$r$ and omnibus CF tests are the most significant, outperforming the Clark-Evans test.
\end{modnew}

\section{Discussion}

\begin{modnew}
    The exact relationship between the choice of the weight function and the power of the CF tests for different alternative hypotheses is still unclear and requires further reasearch. We speculate that if the alternative is a Gibbs process, the power of the test may depend on the similarity between the $\xi$ function and the pairwise interaction function of the Gibbs process. This is at least true for the Strauss process, where theoretically optimal statistical inference can be performed with the empirical $K$-function \citep{baddeley2015spatial}, which includes the Bessel-weighted CF statistic as a variant.
\end{modnew}

The empirical CF has a unified formula even for non-rectangular study regions, e.g., disks, polygons, or disconnected shapes (as in sparsely-sampled regions), in contrast to the empirical CDF. This makes extending the CF test to such scenarios another valuable future research target.

The CF test can also be applied to other types of null hypotheses for spatial point processes, for example, whether a specific point process model (e.g. an inhomogeneous Poisson process with a particular form of intensity function) is appropriate for the point pattern studied. See \citet{JIMENEZGAMERO20093957} for CF tests for scenarios in which the null distribution has unknown parameters.

\begin{modnew}
    The three-dimensional CF statistic may be used to test for space-time CSR, but in practice, testing for space-time interaction is usually of more interest than testing for CSR. Research is currently underway to extend the CF test to such cases as a test of independence.
\end{modnew}

\section*{Appendix}

\begin{modnew}
    The Appendix includes a simulation on the power of the three-dimensional CF test, a simulation on the robustness of the CF tests to boundary points, further mathematical and implementation details on the asymptotic distribution of the CF test, and the proofs of every proposition contained in this paper and the Appendix.
\end{modnew}

\appendix

\begin{modnew}
    \section{Simulation on the power of the three-dimensional CF test}

    Similar to the simulation of the two-dimensional CF test power in Section \ref{chap:power}, we conduct a simulation to examine the power of the CF test when the study region is the cube $[0, 1]^3$. We try to keep the simulation settings the same as the two-dimensional case, but because none of the three simulation settings are implemented in the \texttt{spatstat} package, the \texttt{stpp} package is used instead to simulate the Mat\'ern cluster process and inhomogeneous Poisson process. As for the regularity scenario, due to the lack of existing implementation, the Diggle-Gates-Stibbard process is replaced with a type of sequential inhibition process that is similar in pairwise interaction. Specifically, we define the same function $c(r) = \sin^2\{(\pi/2)r/\rho\}$ for $r < \rho$ and $c(r) = 1$ for $r \ge \rho$, but generate uniformly-distributed points sequentially, each having accepting probability $c(r)$, where $r$ is the distance to the closest already generated point. The $L$-test is included alongside the CF test for comparison. It is defined the same way as the two-dimensional case, except that now the parameter $s = 1.25/n^{1/3}$.

    The simulation result is presented in Figure \ref{fig:power3d}. The sample size is still chosen as $n = 25, 100, 1000$. The size of the tests is $\alpha = 0.05$ and the corresponding (two-sided) thresholds under the fixed $n$’s are estimated beforehand via Monte Carlo simulations of 5000 CSR patterns. The power of each test is then estimated from 2000 independent simulated realizations for each alternative scenario. The simulated Mat\'ern cluster process has the same set of cluster radii as the two-dimensional case, $\delta \in \{0.075, 0.15, 0.3\}$. The intensity function of the inhomogeneous Poisson process has been naturally extended to
    \[ \lambda(x_1, x_2, x_3) = \{\theta_1 - (\theta_1-1)x_1\}\{\theta_2 - (\theta_2-1)x_2\}\{\theta_3 - (\theta_3-1)x_3\}, \]
    where $(\theta_1, \theta_2, \theta_3) = (1, 1, \theta_{\text{mid}}), (\theta_{\text{mid}}, \theta_{\text{mid}}, \theta_{\text{mid}}), (\theta_{\text{mid}}, \theta_{\text{high}}, \theta_{\text{high}})$ for scenario index I, II and III, respectively. The parameters $\theta_{\text{mid}} = 1.5, 1.2, 1.1$ and $\theta_{\text{high}} = 3, 1.5, 1.2$ for $n = 25, 100, 1000$. The maximum inhibition distance $\rho$ of the sequential inhibition process is shown in the figure.

    As shown in Figure \ref{fig:power3d}, the Bonferroni-corrected omnibus CF test has even better margins over the $L$-test in most three-dimensional scenarios than it does when $D = 2$. Its advantage is especially pronounced in the settings of clustering, heterogeneity and large sample sizes.

    \begin{figure}
        \figuresize{.8}
    \figurebox{20pc}{25pc}{}[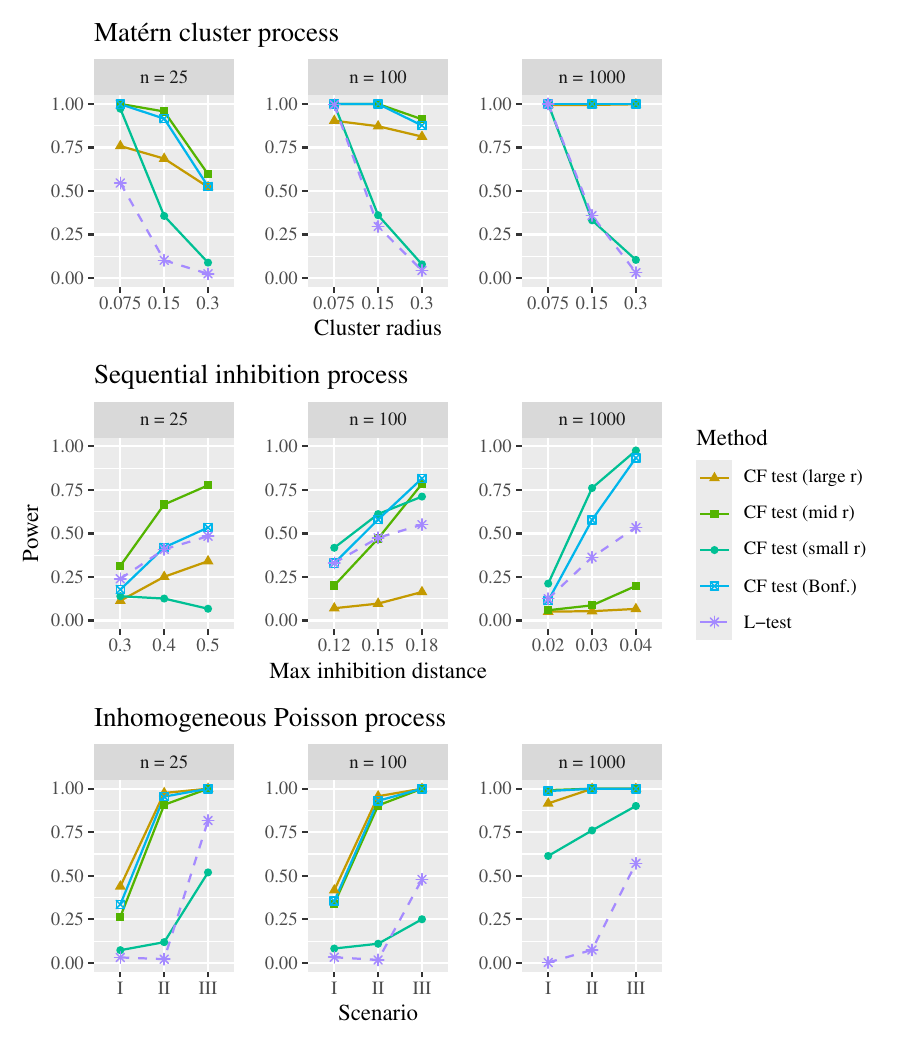]
        \caption{Simulation results of the three-dimensional CSR test power under different alternative hypotheses, comparing the CF test (solid lines) with the $L$-test (dashed lines).}
        \label{fig:power3d}
    \end{figure}
    
\end{modnew}

\begin{modnew}
    \section{Simulation on the robustness of the CF tests to boundary points}

    Theoretically, the CF test should not be sensitive to points on the boundary. Here, we provide a simulation that corroborates this statement. The study region $[0,1]^2$ of the first four example point patterns shown in Figure \ref{fig:datasets} are `eroded' by a distance $\delta$, where $\delta = 0, 0.002, \ldots, 0.04$. The eroded study region is $[\delta, 1 - \delta]^2$. For each $\delta$, we only test the points that are inside the eroded study region and derive the $p$-values via Monte Carlo simulation of $5,000$ CSR patterns. The CF tests we use are the same as those in Section \ref{chap:simulation}. The results are shown in Figure \ref{fig:erosion}. As expected, changing the erosion distance does not significantly alter the results of the CF tests. Only the $p$-values of the Bonferroni-corrected CF test for the scouring rushes pattern are slightly more variable, probably because of its relatively smaller sample size. 

    \begin{figure}[h]
        \figuresize{.7}
        \figurebox{20pc}{25pc}{}[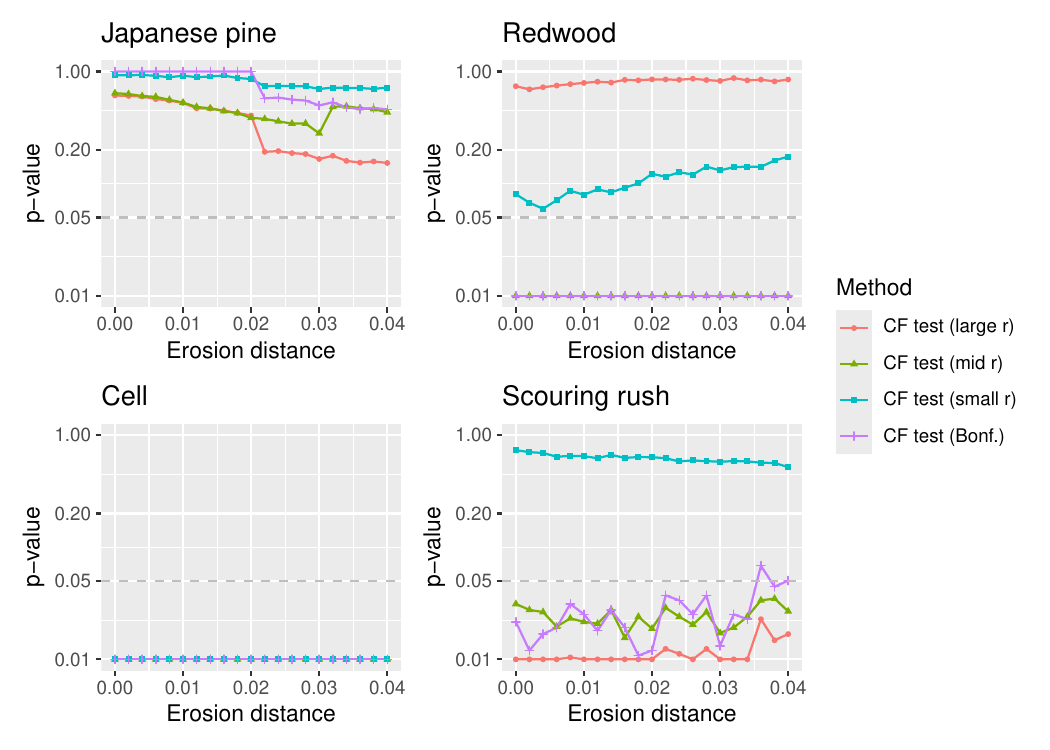]
        \caption{The $p$-value of the CF tests versus the erosion distance $\delta$ for the four example point patterns. The gray dashed line represents the typical significance level $0.05$.}
        \label{fig:erosion}
    \end{figure}
\end{modnew}

\section{Further details \begin{modnew}
    on the asymptotic distribution of the Cauchy-weighted CF test statistic
\end{modnew}}

\begin{modnew}

\subsection{Main procedure}

Finding the eigenvalues of the linear operator $T$ defined by Equation (5) of the main text is in general a very difficult task, but
we will show here that a special matrix form exists under a certain basis of $\mathcal{L}^2(\mathcal{A})$,
when using independent Cauchy weights and the study region $\mathcal{A} = [0, 1]^D$, such that the eigenvalues and asymptotic distribution can be found using a very efficient algorithm.

    Because $T$ is compact and self-adjoint, most of the usual finite-dimensional linear algebra results still apply. Under a certain basis, write the (infinite) matrix form for $T$ as $\bm{T}$, then
\begin{equation}
    \label{eq:T_D_dim_decomposition}
    \bm{T} = \bm{A}^{(D)} - \bm{B}^{(D)} - \bm{B}^{(D)\intercal} + \bm{C}^{(D)},
\end{equation}
where each matrix on the right hand side corresponds to a term at the same position in the kernel of $T$ in Equation \eqref{eq:linear_operator}. Since $\xi$ is the joint CF of independent Cauchy variables, it is separable. Therefore, the matrices in Equation \eqref{eq:T_D_dim_decomposition} are the $D$th Kronecker power of the one-dimensional counterparts,
\begin{gather}
    \bm{T} = \bm{A}^{\otimes D} - \bm{B}^{\otimes D} - \bm{B}^{\intercal\otimes D} + \bm{C}^{\otimes D}, \label{eq:tdecomp}
\end{gather}
where $\bm{A}$, $\bm{B}$ and $\bm{C}$ are respectively equal to $\bm{A}^{(1)}$, $\bm{B}^{(1)}$ and $\bm{C}^{(1)}$, the decomposition of $\bm{T}$ when dimension $D=1$. Now our task is to find these three matrices.

Select an orthonormal basis for $\mathcal{L}^2([0, 1])$ to be $\{\,1,\allowbreak \; 2^{1/2} \cos 2k\pi x, \allowbreak \; 2^{1/2} \sin 2k\pi x \mid k = 1, 2, \ldots \,\}$. The matrices $\bm{A}$, $\bm{B}$ and $\bm{C}$ are then obtained via inner products with the basis functions. The results have nice closed forms. They also conveniently split into orthogonal cosine (including constants) and sine parts (or `even' and `odd' parts if $1/2$ is considered to be the center),
$$\bm{A} = \bm{A}_1 \oplus \bm{A}_2,\quad
    \bm{B} = \bm{B}_1 \oplus \bm{0},\quad \bm{C} = \bm{C}_1 \oplus \bm{0},
    $$
where the direct sum $\oplus$ produces a diagonal block matrix containing the addends, i.e., $\bm{A}_1 \oplus \bm{A}_2 = \diag(\bm{A}_1, \bm{A}_2)$. The actual values of the matrices are
\begin{gather*}
    \bm{A}_1 = \begin{bmatrix}
        \alpha, & -\beta\bm{u}\\
        -\beta\bm{u}, & \bm{U} - (1-e^{-\rho})\bm{u}\bm{u}^\intercal
    \end{bmatrix}, \quad \bm{A}_2 = \bm{U} + (1-e^{-\rho}) \bm{v}\bm{v}^\intercal, \\
    \bm{B}_1 = \begin{bmatrix}
        \alpha & \bm{0}^\intercal\\
        -\beta\bm{u} & \bm{0}
    \end{bmatrix},\quad \bm{C}_1 = \begin{bmatrix}
        \alpha & \bm{0}^\intercal\\
        \bm{0} & \bm{0}
    \end{bmatrix};\\
    \bm{U} = \diag(\bm{u}), \quad
    u_j = \frac{2\rho}{(2\pi j)^2 + \rho^2},\quad v_j = \frac{2\pi j}{\rho}u_j,\quad j = 1, 2, \ldots, \\
    \alpha = \frac{2(e^{-\rho}+\rho-1)}{\rho^2}, \quad \beta = \frac{\sqrt{2}(1-e^{-\rho})}{\rho}.
\end{gather*}
Here, we use the substitution $\rho = 1/r$ to avoid fractions in the exponential.

Because the sine parts of $\bm{B}$ and $\bm{C}$ are zero, under a certain permutation of the $D$th tensor power of the one-dimensional basis, the binomial expansion gives
\begin{gather}
    \bm{T} = (\bm{A}_1^{\otimes D} - \bm{B}_1^{\otimes D}- \bm{B}_1^{\intercal\otimes D} + \bm{C}_1^{\otimes D}) \oplus \bigoplus_{d=1}^D \bm{I}_{\binom{D}{d}} \otimes (\bm{A}_1^{\otimes (D-d)} \otimes \bm{A}_2^{\otimes d}).
    \label{eq:T_binomial_expansion}
\end{gather}
This means that the eigenvalues of $\bm{T}$ can be easily derived from those of of $\bm{A}_1$, $\bm{A}_2$, and $\bm{A}_1^{\otimes D} - \bm{B}_1^{\otimes D}- \bm{B}_1^{\intercal\otimes D} + \bm{C}_1^{\otimes D}$: The eigenvalues of a Kronecker product are all the possible products between the eigenvalues of its factors, and those of a direct sum are the union of the addends' eigenvalues.

\begin{proposition} 
    \label{prop_one_dim_A_eigenvalues}
    The eigenvalues $\lambda =  2\rho / (\tau^2 + \rho^2)$ of $\bm{A}_1$ and $\bm{A}_2$ are solutions to the two equations
    \begin{gather*}
        \tan \frac{\tau}{2} = \frac{\rho}{\tau},\quad
        \tan \frac{\tau}{2} = -\frac{\tau}{\rho}, \quad
        \tau > 0,
    \end{gather*}
    respectively.
\end{proposition}
A proof is given in Section \ref{proof_prop_one_dim_A_eigenvalues}.

The two equations can be written in the numerically more stable forms $\tau \sin\frac{\tau}{2} - \rho\cos\frac{\tau}{2} = 0$ and $\rho \sin\frac{\tau}{2} + \tau\cos\frac{\tau}{2} = 0$.
There is one and only one solution $\tau_k$ in every interval $((2k - 2)\pi, (2k-1)\pi) \ (k = 1, 2, \ldots)$ for the first equation, and in every interval $((2k-1)\pi, 2k\pi)$ for the second equation. Both will have roots becoming increasingly close to the left boundary as $k$ grows. We employ Halley's root-finding method, safeguarded by bisection, as implemented in the C++ Boost package.

As for the matrix $\bm{S}_D=\bm{A}_1^{\otimes D} - \bm{B}_1^{\otimes D}- \bm{B}_1^{\intercal\otimes D} + \bm{C}_1^{\otimes D}$, we have the following result.
\begin{proposition}
    \label{prop_S_eigenvalues}
    Let $\lambda_k$ be the $k$th largest eigenvalue of $\bm{A}_1^{\otimes D}$, of the form $\lambda_{k_1}^{(1)}\ldots\lambda_{k_D}^{(1)}$, where $\lambda_{k_d}^{(1)}$ is the $k_d$th eigenvalue of $\bm{A}_1$. Then the $k$th largest eigenvalue $\mu_k$ of $\bm{S}_D$ belongs to the interval $[\lambda_{k+1}, \lambda_{k}]$. If $\lambda_{k+1} = \lambda_{k}$, then $\mu_k = \lambda_{k}$. Otherwise, $\mu_k \in (\lambda_{k+1}, \lambda_{k})$ and is the unique solution (inside the interval) to the equation
    \begin{equation}
        \label{eq:rank_two_update}
        \left\{1 - F_1(\mu_k)/\alpha^D\right\}^2 + F_0(\mu_k) \left\{1 - F_2(\mu_k)/\alpha^{2D}\right\} = 0,
    \end{equation}
    where
    \begin{gather*}
        F_\ell(\mu) = \sum_{j=1}^\infty \frac{\lambda_j^{\ell}\zeta^{2}_{j}}{\lambda_j - \mu}, \quad \ell = 0, 1, 2; \quad \zeta_{k} = \prod_{d=1}^D \frac{2\alpha^{1/2} \lambda_{k_d}^{(1)}}{\{(\lambda_{k_d}^{(1)} + 1)(2 - \lambda_{k_d}^{(1)}\rho)\}^{1/2}}.
    \end{gather*}
\end{proposition}
A proof is given in Section \ref{proof_prop_S_eigenvalues}.

To solve the equation, it is important to know how close $\mu_k$ is to either of its boundaries. We have 
\[ \sum_{k = 1}^\infty (\lambda_k - \mu_k) = \mathrm{tr}(\bm{A}_1^{\otimes D}) - \mathrm{tr}(\bm{S}_D) = \alpha^D, \quad \sum_{k = 1}^\infty (\mu_k - \lambda_{k+1}) = \lambda_1 - \alpha^D. \]
Both are sums of positive differences. The latter is closer to zero, especially when $\rho$ is large ($O(1/\rho^2)$ compared to $O(1/\rho)$ for the former), so the solution $\mu_k$ should be close to the smaller eigenvalue $\lambda_{k+1}$. This is also confirmed directly using numerical methods.

Matrix $\bm{S}_D$ is a symmetric rank-two update of $\bm{A}_1^{\otimes D}$, as shown in the proof of Proposition \ref{prop_S_eigenvalues} in Section \ref{proof_prop_S_eigenvalues}. \citet{Bunch1978} and \citet{li1994} developed techniques for solving the characteristic equation of a rank-one update, but the case of a rank-two update is much more complicated. Nonetheless, we can still borrow their ideas and approximate $F_\ell(\mu)$ by a rational function
\begin{equation}
    \label{eq:rankone_approximation}
    F_\ell(\mu_k) \approx \frac{a_\ell}{\lambda_{k} - \mu_k} + b_\ell + C_{k+1}\frac{\lambda_{k+1}^{\ell}\zeta^{2}_{k+1}}{\lambda_{k+1} - \mu_k},
\end{equation}
where $C_{k+1}$ is the multiplicity of $\lambda_{k+1}$. This is the fixed weight variant of \citet{li1994}. It preserves the $(k+1)$th term in its exact form, because $\mu_k$ is the closest to $\lambda_{k+1}$ and thus that term needs to be as accurate as possible. A constraint $a_1^2=a_0a_2$ is imposed so that no quadratic term $1/(\lambda_k - \mu_k)^2$ exists. Otherwise, the approximation can be very inaccurate close to $\lambda_k$, since the original secular equation contains no such term.

Applying the approximation in Equation \eqref{eq:rankone_approximation} to Equation \eqref{eq:rank_two_update}, the equation that we aim to solve, ultimately leads to an approximate quadratic equation for $\mu_k$,
\begin{equation}
    \label{eq:quadratic_approximation}
    s_2(\lambda_{k} - \mu_k)(\lambda_{k+1} - \mu_k) + s_1(\lambda_{k+1} - \mu_k) - s_4 C_{k+1}\zeta^{2}_{k+1}(\lambda_{k} - \mu_k) - s_3 C_{k+1}\zeta^{2}_{k+1} = 0,
\end{equation}
See Section \ref{deriv_quadratic_approximation} for how to derive and solve this equation, and the values of $a_\ell$, $b_\ell$ and the equation coefficients. Starting at a plausible guess for $\mu_k$, this approximate equation is solved repeatedly by an iterative algorithm. At iteration $t$, the coefficients in the approximation are updated using the new approximate solution $\mu_k^{(t)}$ for $\mu_k$, so that the approximation agrees with the original equation at $\mu_k^{(t)}$. Usually, one or two iterates are sufficient to arrive at an accurate value for the true solution $\mu_k$ to Equation \eqref{eq:rank_two_update}.

Now we have found all the eigenvalues of $\bm{A}_1$, $\bm{A}_2$, and $\bm{S}_D$. The eigenvalues of $\bm{T}$, according to Equation \eqref{eq:T_binomial_expansion}, are either $\lambda_j^{\bm{S}_D}$ or of the form $\lambda_{j_1}^{\bm{A}_1} \cdots \lambda_{j_{D-d}}^{\bm{A}_1} \lambda_{j_{D-d+1}}^{\bm{A}_2} \cdots \lambda_{j_D}^{\bm{A}_2}$ with a multiplicity of $\binom{D}{d}$, for $d = 1, \ldots, D$. Here, the superscripts denote the matrix to which an eigenvalue corresponds. For example, when the dimension $D = 2$, the eigenvalues are $\lambda_j^{\bm{S}_2}$, $\lambda_{j_1}^{\bm{A}_2}\lambda_{j_2}^{\bm{A}_2}$, or $\lambda_{j_1}^{\bm{A}_1}\lambda_{j_2}^{\bm{A}_2}$ with multiplicity 2.

Write the eigenvalues we have obtained as a decreasing sequence $(\lambda_j)_{j=1}^\infty$. \citet{imhof} found a formula to calculate the CDF of a weighted sum of squared standard normal variables, which is the asymptotic distribution of our test statistic $\Delta$. This formula yields
\begin{equation}
    \label{eq:imhof}
    \mathrm{pr}(\Delta \le x)|_{n \rightarrow \infty} = \frac{1}{2} - \frac{1}{\pi}\int_0^\infty \frac{\sin\{\theta(u) - \frac{1}{2} x u\}}{u}e^{-\eta(u)}\,du,
\end{equation}
where 
\begin{gather*}
    \theta(u) = \frac{1}{2}\sum_{j = 1}^\infty \arctan (\lambda_j u), \quad
    \eta(u) = \frac{1}{4}\sum_{j = 1}^\infty \log(1+\lambda_j^2 u^2).
\end{gather*}
In the original procedure, the summation is over a finite set of $\lambda_j$, but the proof provided by \citet{imhof} directly extends to infinite summations, which is what we use here. See Section \ref{deriv_imhof} for how to evaluate the functions $\theta(u)$ and $\eta(u)$. The integration step is achieved numerically using the Gauss-Konrod quadrature algorithm from the C++ Boost package. 

The Imhof procedure finds the CDF value at a given point. It can give the $p$-values of a CF test directly, but not the test thresholds, which requires the quantile function. We find the quantile function by inverting the CDF, using the `bracket and solve' root-finding method, again from the C++ Boost package. In this algorithm, the quantile of the log-normal distribution with the same mean and variance can serve as a quite effective initial guess.

\end{modnew}

\subsection{Quadratic approximation of the rank-two update characteristic equation} \label{deriv_quadratic_approximation}

Here, we derive and present the parameters for the quadratic approximation in Equation \eqref{eq:quadratic_approximation}.  \begin{modnew}
    Starting from a
\end{modnew} guess $\mu_0$,  we can approximate $F_\ell(\mu)$ based on its behavior around $\mu_0$. The approximation we choose is the rational function in Equation \eqref{eq:rankone_approximation},
\begin{equation*}
    \hat{F}_\ell(\mu) = \frac{a_\ell}{\lambda_{k} - \mu} + b_\ell + C_{k+1}\frac{\lambda_{k+1}^{\ell}\zeta^{2}_{k+1}}{\lambda_{k+1} - \mu}.
\end{equation*}
Ideally, the parameters $a_\ell$ and $b_\ell$ are chosen so that the approximant function agrees with the original function in value and derivative at $\mu_0$. That is,  \begin{modnew}
    so that
\end{modnew}
\begin{gather*}
    F_\ell(\mu_0) = \hat{F}_\ell(\mu_0) = \frac{a_\ell}{\lambda_{k} - \mu_0} + b_\ell + C_{k+1}\frac{\lambda_{k+1}^{\ell}\zeta^{2}_{k+1}}{\lambda_{k+1} - \mu_0},\\
    F'_\ell(\mu_0) = \hat{F}_\ell'(\mu_0) = \frac{a_\ell}{(\lambda_{k} - \mu_0)^2} + C_{k+1}\frac{\lambda_{k+1}^{\ell}\zeta^{2}_{k+1}}{(\lambda_{k+1} - \mu_0)^2}.
\end{gather*} However, due to the added constraint $a_1^2=a_0a_2$, the number of equations exceeds that of variables. There are no solutions. Therefore, we instead only require the derivatives at $\mu_0$ to approximately agree. To elaborate, notice that the constraint $a_1^2=a_0a_2$ is equivalent to $\log(a_\ell)$ being a linear function of $\ell$. Let $\tilde{a}_\ell$ be the exact solutions to $F'_\ell(\mu_0) = \hat{F}_\ell'(\mu_0)$,
\[ 
    \tilde{a}_\ell = (\lambda_{k} - \mu_0)^2\left\{F'_\ell(\mu_0) - C_{k+1}\frac{\lambda_{k+1}^{\ell}\zeta^{2}_{k+1}}{(\lambda_{k+1} - \mu_0)^2}\right\},
\]
which can then be used to form a linear regression problem, i.e., to regress $\log(\tilde{a}_\ell)$ on $\ell$. The fitted values are the solutions $a_\ell$ we desire,
\[ 
    a_1 = (\tilde{a}_1^5 \tilde{a}_2^2 \tilde{a}_3^{-1})^{1/6}, \quad a_2 = (\tilde{a}_1 \tilde{a}_2 \tilde{a}_3)^{1/3}, \quad a_3 = (\tilde{a}_1^{-1} \tilde{a}_2^2 \tilde{a}_3^5)^{1/6}.
\]

For parameter $b_\ell$, solving $F_\ell(\mu_0) = \hat{F}_\ell(\mu_0)$ gives 
\[ b_\ell = F_\ell(\mu_0) - \frac{a_\ell}{\lambda_{k} - \mu_0} - C_{k+1}\frac{\lambda_{k+1}^{\ell}\zeta^{2}_{k+1}}{\lambda_{k+1} - \mu_0}. \]

By substituting $F_\ell(\mu)$ with its approximation $\hat{F}_\ell(\mu)$ in Equation \eqref{eq:rank_two_update}, we obtain an approximate equation,
\begin{gather*}
    s_2 + \frac{s_1}{\lambda_{k} - \mu_k} - \frac{C_{k+1}\zeta'^{2}_{k+1} s_4}{\lambda_{k+1} - \mu_k} - \frac{C_{k+1}\zeta'^{2}_{k+1} s_3}{(\lambda_{k} - \mu_k)(\lambda_{k+1} - \mu_k)} = 0,
\end{gather*}
where
\begin{gather*}
    s_1 = 2 a_1 b_1 - a_0b_2 - a_2b_0 - 2a_1\alpha^D+a_0\alpha^{2D},\\
    s_2 = b_1^2 - b_0b_2 - 2b_1\alpha^D+(b_0 + 1)\alpha^{2D},\\
    s_3 = a_0 \lambda^{2}_{k+1} -2a_1\lambda_{k+1}+a_2, \\
    s_4 = b_0 \lambda^{2}_{k+1} -2b_1\lambda_{k+1}+b_2 - \alpha^{2D} + 2\alpha^D \lambda_{k+1}.
\end{gather*}
It can be transformed into a quadratic equation with at most two roots. Most of the time, it has only one root inside $(\lambda_k, \lambda_{k+1})$, the same as the original characteristic equation. Occasionally, both roots are inside the interval. Then the one closer to $\lambda_{k+1}$ is used, because the approximation is more accurate there. For large $k$ and small $\rho$, it may happen that no roots can be found, in which case $\mu_k$ is simply replaced by $\lambda_{k+1}$ since they are close at such $k$ anyway.

Additionally, when evaluating $F_\ell(\mu_0)$ and $F_\ell'(\mu_0)$, because there are infinite terms in the summation, only the first $J$ terms in $F_\ell$ are summed up. The remainder of the series in $F_\ell(\mu)$ is, using the Taylor expansion at $\lambda_{j} = 0$ assuming $\mu > \lambda_{j}$ for $j>J$,
\begin{align*}
    F_{\ell, -J}(\mu) &= \sum_{j = J+1}^\infty \frac{\lambda_j^{\ell}\zeta'^{2}_{j}}{\lambda_{j} - \mu}\\ 
    &= -\sum_{r=0}^\infty \frac{1}{\mu^{r+1}} \sum_{j = J+1}^\infty \lambda^{\ell+r}_{j}\zeta'^{2}_{j}\\
    &= -\sum_{r=0}^\infty \frac{1}{\mu^{r+1}} \left[\left(\sum_{k = 1}^\infty \lambda^{\begin{modnew}(1)\end{modnew}\ell+r}_{k}\zeta'^{\begin{modnew}(1)\end{modnew}2}_{k}\right)^D - \sum_{j = 1}^J \lambda^{\ell+r}_{j}\zeta'^{2}_{j}\right].
\end{align*}
Simply taking its derivative will give the remainder for $F'_\ell(\mu)$. Each term in the series is less than $\lambda_{N+1}/\mu$ times the preceding term, so higher order terms can be safely discarded. The coefficients in the equation can be pre-calculated using the identity (see Section \ref{proof_prop_S_eigenvalues})
\begin{gather*}
    \sum_{j=1}^\infty \lambda_j^{\begin{modnew}(1)\end{modnew}m} \zeta'^{\begin{modnew}(1)\end{modnew}2}_j = \bm{\zeta}'^\intercal \bm{\Lambda}^m\bm{\zeta}' = \bm{\zeta}^\intercal \bm{A}_1^m\bm{\zeta} = \alpha (\bm{A}_1^m)_{11}.
\end{gather*}
Although we have found no easy algorithm for $(\bm{A}_1^m)_{11}$, the lower degree ones can be calculated in reasonable closed forms
\begin{align*}
    (\bm{A}_1^0)_{11} &= 1, \quad (\bm{A}_1^1)_{11} = \alpha, \quad (\bm{A}_1^2)_{11} = \alpha ^2+\beta ^2 G_2,\\
    (\bm{A}_1^3)_{11} &= \alpha ^3+2 \alpha  \beta ^2 G_2-\beta ^2 \gamma  G_2^2+\beta ^2 G_3,\\
    (\bm{A}_1^4)_{11} &= \alpha ^4+3 \alpha ^2 \beta ^2 G_2-2 \alpha  \beta ^2 \gamma  G_2^2+2 \alpha  \beta ^2 G_3+\beta ^4 G_2^2+\beta ^2 \gamma ^2 G_2^3\\
    &\quad\;-2 \beta ^2 \gamma  G_2 G_3+\beta ^2 G_4,\\
    (\bm{A}_1^5)_{11} &= \alpha ^5+4 \alpha ^3 \beta ^2 G_2-3 \alpha ^2 \beta ^2 \gamma  G_2^2+3 \alpha  \beta ^4 G_2^2+2 \alpha  \beta ^2 \gamma ^2 G_2^3-2 \beta ^4 \gamma  G_2^3\\
    &\quad\;+3 \alpha ^2 \beta ^2 G_3-4 \alpha  \beta ^2 \gamma  G_2 G_3+2 \beta ^4 G_2 G_3-\beta ^2 \gamma ^3 G_2^4+3 \beta ^2 \gamma ^2 G_2^2 G_3\\
    &\quad\;+2 \alpha  \beta ^2 G_4-2 \beta ^2 \gamma  G_2 G_4-\beta ^2 \gamma  G_3^2+\beta ^2 G_5,
\end{align*}
where $\gamma = 1 - e^{-\rho}$ and
\begin{align*}
    G_m &= \sum_{j=1}^\infty \left\{\frac{2\rho}{(2\pi j)^2 + \rho^2}\right\}^m, \\
    G_2 &= \frac{\rho ^2+\rho  \sinh (\rho )-4 \cosh (\rho )+4}{2 \rho ^2 \{\cosh (\rho )-1\}},\\
    G_3 &= \frac{\rho ^3 \coth \left(\rho / 2\right)+3 \rho ^2+3 \rho  \sinh (\rho )-16 \cosh (\rho )+16}{4 \rho ^3 \{\cosh (\rho )-1\}},\\
    G_4 &= \frac{e^{2 \rho }}{3 \left(e^{\rho }-1\right)^4 \rho ^4} [2 \rho ^4-15 \rho ^2+3 \rho  \sinh (\rho ) \{2 \rho ^2+5 \cosh (\rho )-5\}\\
    &\quad\;+(\rho ^4+15 \rho ^2+192) \cosh (\rho )-48 \cosh (2 \rho )-144],\\
    G_5 &= -\frac{16}{\rho^5} + \frac{e^{5 \rho / 2}}{24\rho^4\left(e^{\rho }-1\right)^5}\biggl[20 \rho  \sinh \left(\frac{\rho }{2}\right) \{4 \rho ^2+(2 \rho ^2+21) \cosh (\rho )-21\}\\
    &\quad\;+(22 \rho ^4-90 \rho ^2+210) \cosh \left(\frac{\rho }{2}\right)+(2 \rho ^4+90 \rho ^2-315) \cosh \left(\frac{3 \rho }{2}\right)\\
    &\quad\;+105 \cosh \left(\frac{5 \rho }{2}\right)\biggr].
\end{align*}

\subsection{Evaluation of functions $\theta(u)$ and $\eta(u)$ in Imhof's procedure} \label{deriv_imhof}

Functions $\theta(u)$ and $\eta(u)$ are determined by formulas
\begin{gather*}
    \theta(u) = \frac{1}{2}\sum_{j=1}^\infty \arctan (\lambda_j u), \quad
    \eta(u) = \frac{1}{4}\sum_{j=1}^\infty \log(1+\lambda_j^2 u^2).
\end{gather*}
They require infinitely many eigenvalues, but we can only obtain finitely many, so the functions cannot be calculated directly. Suppose we have obtained the first $J$ eigenvalues, $\lambda_j\ (j = 1, \ldots, J)$, where $J$ is sufficiently large. The terms in $\theta(u)$ and $\eta(u)$ corresponding to $j = 1, \ldots, J$ can be explicitly evaluated. Let the remainders of the two series be 
\[ \theta_{-J}(u) = \frac{1}{2}\sum_{j=J+1}^\infty \arctan (\lambda_j u), \quad  \eta_{-J}(u) = \frac{1}{4}\sum_{j=J+1}^\infty \log(1+\lambda_j^2 u^2).\]
Then since $\lambda_j$ should be close to zero for $j > J$, the remainders can be approximated by the second-order Taylor expansion of each term at $\lambda_j u = 0$,
\begin{gather*}
    \theta_{-J}(u) \approx \frac{1}{2}u \left(\sum_{j=1}^\infty \lambda_j - \sum_{j=1}^J \lambda_j\right), \quad
    \eta_{-J}(u) \approx \frac{1}{4}u^2\left(\sum_{j=1}^\infty \lambda_j^2 - \sum_{j=1}^J \lambda_j^2\right),
\end{gather*}
where
\[ \sum_{j=1}^\infty \lambda_j = E (\Delta), \quad \sum_{j=1}^\infty \lambda_j^2 = \lim_{n\rightarrow\infty} \mathrm{var} (\Delta), \]
a result of
\[ \Delta \rightarrow \sum_{j=1}^\infty \lambda_j Z^2_j \]
in distribution. The moments are given in Section \ref{chap:first_two_moments}.
The Taylor expansion is accurate as long as $\lambda_{J+1}u$ is not too large. For large $u$, the approximation is off but not dangerously so, because large $u$ would shrink the integrand in Equation \eqref{eq:imhof} and force it to have a stronger Gaussian rate of decay.

Even though only the first $J$ terms are explicitly computed, the summation 
\[ \sum_{j=1}^J \arctan (\lambda_j u) \]
in function $\theta(u)$ still involves many expensive evaluations of the arctangent function. However, the majority of them are avoidable, because there is a nice bound for the relative error of the first-order Taylor expansion of $\arctan\left(x+\epsilon\right)$,
\begin{gather*}
    \left|\frac{\arctan\left(x+\epsilon\right) - \arctan x - \frac{\epsilon}{1+x^2}}{\arctan\left(x+\epsilon\right)}\right| < 10^{-6},
\end{gather*}
for $|\epsilon| < 10^{-3}$. So, when evaluating $\arctan (\lambda_j u)$, we can safely replace $\arctan (\lambda_j u)$ with the expansion $\arctan(\lambda'_ju) + (\lambda_j - \lambda'_j)u/(1+\lambda_j'^2u^2)$ for $j' < j$ and $\lambda'_j - \lambda_j < 10^{-3}$, with a relative error less than $10^{-6}$. Simulation shows that least 95\% of the evaluations of the arctangent function can be replaced this way for various $\rho$. 

As for the finite summation in function $\eta(u)$, it is more efficient to evaluate its exponential
$$\exp\left\{\sum_{j=1}^J \log(1+\lambda_j^2 u^2)\right\} = \prod_{j=1}^J (1+\lambda_j^2 u^2).$$
However, this form can be numerically unstable due to many factors containing small $\lambda_j u$, so such factors are grouped together and approximated using the first-order Taylor expansion
\[ \prod_j (1+\lambda_j^2 u^2) \approx 1 + \sum_j \lambda_j^2 u^2. \]

\section{Proofs of some propositions}

\subsection{Proof of Proposition~\ref{prop_first_two_moments}}
To obtain the mean and variance of the test statistic under the null hypothesis, observe that
\begin{align*}
    E_0(\Delta) &= \int_{\mathbb{R}^D}nE_0\left\{\left|\hat{\phi}_X(\bm{t})-\phi(\bm{t})\right|^2\right\} w(\bm{t}) d\bm{t}\\
    &= \int_{\mathbb{R}^D}\frac{1}{n} \sum_{j, k = 1}^n E_0 \left[\left\{e^{i\bm{t}^\intercal\bm{x}_j} - E_0 \left(e^{i\bm{t}^\intercal\bm{x}_j}\right)\right\}\left\{e^{-i\bm{t}^\intercal\bm{x}_k} - E_0 \left(e^{-i\bm{t}^\intercal\bm{x}_k}\right)\right\}\right] w(\bm{t}) d\bm{t}\\
    &= \int_{\mathbb{R}^D}\frac{1}{n} \sum_{j = 1}^n E_0 \left[\left\{e^{i\bm{t}^\intercal\bm{x}_j} - E_0 \left(e^{i\bm{t}^\intercal\bm{x}_j}\right)\right\}\left\{e^{-i\bm{t}^\intercal\bm{x}_j} - E_0 \left(e^{-i\bm{t}^\intercal\bm{x}_j}\right)\right\}\right] w(\bm{t}) d\bm{t}\\
    &= \int_{\mathbb{R}^D}E_0 \left[\left\{e^{i\bm{t}^\intercal\bm{x}_1} - E_0 \left(e^{i\bm{t}^\intercal\bm{x}_1}\right)\right\}\left\{e^{-i\bm{t}^\intercal\bm{x}_1} - E_0 \left(e^{-i\bm{t}^\intercal\bm{x}_1}\right)\right\}\right] w(\bm{t}) d\bm{t}\\
    &= E_0 \int_{\mathbb{R}^D} \left(1 - e^{i\bm{t}^\intercal\bm{x}_1}e^{-i\bm{t}^\intercal\bm{x}_2}\right) w(\bm{t}) d\bm{t}\\
    &= 1 - E_0 \left\{\xi(\bm{x}_1 - \bm{x}_2)\right\}.
\end{align*}
Using the same trick for the second moment,
\begin{align*}
    E_0 (\Delta^2) &= \int_{\mathbb{R}^D \times \mathbb{R}^D}n^2E_0 \left\{\left|\hat{\phi}_X(\bm{t}_1)-\phi(\bm{t}_1)\right|^2\left|\hat{\phi}_X(\bm{t}_2)-\phi(\bm{t}_2)\right|^2\right\} w(\bm{t}_1)w(\bm{t}_2) d\bm{t}_1d\bm{t}_2\\
    &= 1 - 2E_0\{\xi(\bm{x}_1 - \bm{x}_2)\} + \frac{3n-6}{n}[E_0\{\xi(\bm{x}_1 - \bm{x}_2)\}]^2\\
    &\quad\,+ \frac{2n-2}{n} E_0\left[\{\xi(\bm{x}_1 - \bm{x}_2)\}^2\right] - \frac{4n-8}{n} E_0 \{\xi(\bm{x}_1 - \bm{x}_3)\xi(\bm{x}_2 - \bm{x}_3)\}.
\end{align*}
Therefore, the variance is
\begin{align*}
    \mathrm{var}_0 (\Delta) &= \frac{2n-6}{n}[E_0\{\xi(\bm{x}_1 - \bm{x}_2)\}]^2 + \frac{2n-2}{n} E_0\left[\{\xi(\bm{x}_1 - \bm{x}_2)\}^2\right]\\
            &\quad\, - \frac{4n-8}{n} E_0 \{\xi(\bm{x}_1 - \bm{x}_3)\xi(\bm{x}_2 - \bm{x}_3)\}.
\end{align*}

\subsection{Proof of Proposition \ref{prop_one_dim_A_eigenvalues}} \label{proof_prop_one_dim_A_eigenvalues}

The matrices
$$\bm{A}_1 = \begin{bmatrix}
        \frac{2}{\rho} & \bm{0}^\intercal\\
        \bm{0} & \bm{U}
    \end{bmatrix} - (1-e^{-\rho})\begin{bmatrix}
        \frac{\sqrt{2}}{\rho} \\ \bm{u}
    \end{bmatrix}\begin{bmatrix}
        \frac{\sqrt{2}}{\rho}, & \bm{u}^\intercal
    \end{bmatrix}$$
and $\bm{A}_2 = \bm{U} + (1-e^{-\rho}) \bm{v}\bm{v}^\intercal$ are both symmetric rank-one updates to a diagonal matrix. Eigenvalue problems of this kind has been studied thoroughly. Their eigenvalues are solutions to the characteristic equations \citep{golub1973},
\begin{gather*}
    \det(\bm{A}_1 - \lambda \bm{I}) \propto 1 - (1-e^{-\rho})\left(\frac{2/\rho^2}{2/\rho - \lambda} + \sum_{j = 1}^\infty \frac{u_j^2}{u_j - \lambda}\right) = 0,\\
    \det(\bm{A}_2 - \lambda \bm{I}) \propto 1 + (1-e^{-\rho}) \sum_{j = 1}^\infty \frac{v_j^2}{u_j - \lambda} = 0.
\end{gather*}

Our case is special because the infinite sums can actually be evaluated analytically. Because $\lambda \in (0, 2/\rho)$, we may substitute $\lambda = 2\rho / (\tau^2 + \rho^2)$ with $\tau>0$ into the previous expressions, i.e.,
\begin{gather*}
    \frac{2/\rho^2}{2/\rho - \lambda} + \sum_{j = 1}^\infty \frac{u_j^2}{u_j - \lambda} = \frac{\rho}{2\tau}\cot \frac{\tau }{2} + \frac{1}{2} \coth \frac{\rho }{2}, \\
    \sum_{j = 1}^\infty \frac{v_j^2}{u_j - \lambda}=\frac{\tau}{2\rho}\cot \frac{\tau }{2} - \frac{1}{2} \coth \frac{\rho }{2}.
\end{gather*}
Therefore, the characteristic equations for $\bm{A}_1$ and $\bm{A}_2$ are simplified to
\begin{gather*}
    \tan \frac{\tau}{2} = \frac{\rho}{\tau},\quad
    \tan \frac{\tau}{2} = -\frac{\tau}{\rho}.
\end{gather*}

\subsection{Proof of Proposition \ref{prop_S_eigenvalues}} \label{proof_prop_S_eigenvalues}

To start, the eigenvector $\bm{q}_k$ of $\bm{A}_1$ corresponding to its $k$th eigenvalue $\lambda_k^{(1)} = 2\rho / (\tau_k^2 + \rho^2)$ can be derived in closed form using the formula of \citet{Bunch1978}. It is
\begin{gather*}
    \bm{q}_k^\intercal = \lambda_k^{(1)}  \left(\frac{4-2 \lambda_k^{(1)}  \rho }{\lambda_k^{(1)} +1}\right)^{1/2} \left(\frac{2^{1/2}}{2-\lambda_k^{(1)}\rho}, \quad \frac{\bm{u}}{\bm{u} - \lambda_k^{(1)}\bm{1}}\right),
\end{gather*}
where the division by vector should be understood as a by-element operation. We have used the identity $\tan (\tau_k/2) = \rho/\tau_k$ in forming this equation. Therefore, matrix $\bm{A}_1$ has the eigendecomposition $\bm{A}_1 = \bm{Q}\bm{\Lambda}\bm{Q}^\intercal$, where $\bm{Q}$ is an orthonormal matrix having $\bm{q}_k$ as columns, and $\bm{\Lambda}$ is a diagonal matrix with $\lambda_k^{(1)}$ as diagonal elements. This will be useful later.

The matrix $\bm{S}_D=\bm{A}_1^{\otimes D} - \bm{B}_1^{\otimes D}- \bm{B}_1^{\intercal\otimes D} + \bm{C}_1^{\otimes D}$ is $\bm{A}_1^{\otimes D}$ with vanishing first row and column (this is a result of the whole kernel integrating to zero w.r.t. either variables), and is thus a compression of $\bm{A}_1^{\otimes D}$. By the Cauchy interlacing theorem, $\mu_k \in [\lambda_{k+1}, \lambda_{k}]$, where $\lambda_k$ and $\mu_k$ are the $k$th largest eigenvalue of $\bm{A}_1^{\otimes D}$ and $\bm{S}_D$, respectively. The value of $\lambda_k$ is already known, since it is a product of the form $\lambda_{k_1}^{(1)}\cdots\lambda_{k_D}^{(1)}$. Although each $\lambda_k^{(1)}$ is distinct, $\lambda_k$ can have a multiplicity of $2!, 3!, \ldots,$ or $D!$ from different permutations of the factors. Therefore, those $\mu_k$ that are sandwiched between $\lambda_k = \lambda_{k+1}$ are forced to be the same as $\lambda_k$. As for other $\mu_k$ where $\lambda_k > \lambda_{k+1}$, they are solutions to a particular equation derived below.

To derive the characteristic equation for $\bm{S}_D$, we need to instead represent it as a symmetric rank-two update to $\bm{A}_1^{\otimes D}$,
$$\bm{S}_D=\bm{A}_1^{\otimes D} - \bm{z}^{\otimes D}\bm{z}^{\intercal\otimes D} + (\bm{z}^{\otimes D} - \bm{\zeta}^{\otimes D})(\bm{z}^{\otimes D} - \bm{\zeta}^{\otimes D})^\intercal,$$
where $\bm{z}^\intercal = (\alpha^{1/2},\  (-\beta/\alpha^{1/2}) \bm{u}^\intercal)$ and $\bm{\zeta}^\intercal = (\alpha^{1/2},\  \bm{0}^\intercal)$.  We can partially diagonalize it using the eigendecomposition for $\bm{A}_1$ obtained earlier,
\begin{gather*}
    \bm{Q}^{\intercal\otimes D}\bm{S}_D\bm{Q}^{\otimes D} = \bm{\Lambda}^{\otimes D} - \bm{z}'^{\otimes D}\bm{z}'^{\intercal\otimes D} + (\bm{z}'^{\otimes D} - \bm{\zeta}'^{\otimes D})(\bm{z}'^{\otimes D} - \bm{\zeta}'^{\otimes D})^\intercal,
\end{gather*}
where $\bm{z}' = \bm{Q}^{\intercal} \bm{z}$ and $\bm{\zeta}' = \bm{Q}^{\intercal} \bm{\zeta}$. The elements of $\bm{z}'$ and $\bm{\zeta}'$ are given by
\begin{gather*}
    z'_k = \frac{\lambda_k^{(1)}}{\alpha}\zeta'^{\begin{modnew}(1)\end{modnew}}_k, \quad
    \zeta'^{\begin{modnew}(1)\end{modnew}}_k = \frac{2\alpha^{1/2} \lambda_k^{(1)}}{\{(\lambda_k^{(1)} + 1)(2 - \lambda_k^{(1)}\rho)\}^{1/2}} \quad (k = 1, 2, \ldots).
\end{gather*}
\begin{modnew}
    The superscript $(\cdot)^{(1)}$ in $\zeta'^{(1)}$ is to distinguish it from the quantity $\zeta'^{(D)}_{j}$ introduced later.
\end{modnew}

The general formula of \citet{ARBENZ198875} can be adapted to give the characteristic equation of this kind of symmetric rank-two update to a diagonal matrix,
\begin{align*}
    \det(\bm{Q}^{\intercal\otimes D}\bm{S}_D\bm{Q}^{\otimes D} - \mu\bm{I}) &\propto 1 - \sum_{j=1}^\infty \frac{\left\{\left(\bm{z}'^{\otimes D}\right)_j\right\}^2 - \left\{\left(\bm{z}'^{\otimes D}\right)_j - \left(\bm{\zeta}'^{\begin{modnew}(1)\end{modnew}\otimes D}\right)_j\right\}^2}{(\bm{\Lambda}^{\otimes D})_{jj} - \mu}\\
    &\quad\,- \frac{1}{2}\sum_{j, k = 1}^\infty \frac{\left\{\left(\bm{z}'^{\otimes D}\right)_j\left(\bm{\zeta}'^{\begin{modnew}(1)\end{modnew}\otimes D}\right)_k - \left(\bm{z}'^{\otimes D}\right)_k\left(\bm{\zeta}'^{\begin{modnew}(1)\end{modnew}\otimes D}\right)_j\right\}^2}{\{(\bm{\Lambda}^{\otimes D})_{jj} - \mu\}\{(\bm{\Lambda}^{\otimes D})_{kk} - \mu\}}\\
    &= \left\{1 - \alpha^{-D} F_1(\mu)\right\}^2 + F_0(\mu) \left\{1 - \alpha^{-2D}F_2(\mu)\right\},
\end{align*}
where
\begin{gather*}
    F_\ell(\mu) = \sum_{j=1}^\infty \frac{\lambda_j^{\ell}\left(\zeta'^{(D)}_{j}\right)^2}{\lambda_j - \mu} \quad (\ell = 0, 1, 2),
\end{gather*}
and $\zeta'^{(D)}_{j}$ is the $j$th element of $\bm{\zeta}'^{\otimes D}$ corresponding to $\lambda_j$.  \begin{modnew}
    Outside of this section,
\end{modnew} the entire superscript $(\cdot)'^{(D)}$ is omitted.

\subsection{Proof of Proposition \ref{asymptotic_cumulant}}

Let $\mu_m'$ and $\mu_m$ be the $m$th noncentral and central moment. The $m$th noncentral moment is
\begin{align*}
    \mu_m' & = E(\Delta^m)\\
    &= E \left[\left\{\int_{\mathbb{R}^D}n\left|\hat{\phi}_X(\bm{t})-\phi(\bm{t})\right|^2 w(\bm{t}) d\bm{t}\right\}^m\right]\\
    &= \sum_{\substack{j_\ell, k_\ell = 1\\ (\ell = 1,\ldots,m)}}^n \frac{1}{n^m} E \left[\prod_{\ell=1}^m \int_{\mathbb{R}^D} \left\{e^{i\bm{t}_\ell^\intercal\bm{x}_{j_\ell}} - E \left(e^{i\bm{t}_\ell^\intercal\bm{x}_{j_\ell}}\right)\right\}\left\{e^{-i\bm{t}_\ell^\intercal\bm{x}_{k_\ell}} - E \left(e^{-i\bm{t}_\ell^\intercal\bm{x}_{k_\ell}}\right)\right\} w(\bm{t}_\ell) d\bm{t}_\ell\right].
\end{align*}
Denote the term inside the summation corresponding to the index sequence $(j_\ell, k_\ell)_{\ell=1}^m$ by $M\{(j_\ell, k_\ell)_{\ell=1}^m\}$, then 
\[ \mu_m' = \sum_{\substack{j_\ell, k_\ell = 1\\ (\ell = 1,\ldots,m)}}^n M\left\{(j_\ell, k_\ell)_{\ell=1}^m\right\} \]
and
\[
    M\left\{(j_\ell, k_\ell)_{\ell=1}^m\right\} = \frac{1}{n^m} E \left[\prod_{\ell=1}^m \left\{\xi(\bm{x}_{j_\ell} - \bm{x}_{k_\ell}) - \xi(\bm{y}_\ell - \bm{x}_{k_\ell}) - \xi(\bm{x}_{j_\ell} - \bm{y}_\ell) + \xi(\bm{y}_{\ell} - \bm{y}_{\ell}')\right\}\right],
\]
where $\bm{y}_\ell$ and $\bm{y}'_\ell$ are also mutually independent uniform random variables in $\mathcal{A}$. In the following derivation, we assume $m \ge 2$.

Let us first focus on the case where $j_\ell \neq k_\ell$ for all $\ell$. The term $M\{(j_\ell, k_\ell)_{\ell=1}^m\}$ can be further expanded into a sum of the expectation of some products. Every product has $m$ factors, each being the result of the function $\xi$ applied to the difference of a pair of random vectors. When the whole product contains two unique random vectors, its expectation is asymptotically similar to \begin{modnew}
        $r^D$
    \end{modnew}, otherwise its expectation is asymptotically less than \begin{modnew}
        $r^D$
    \end{modnew},
\begin{gather*}
    E [\{\xi(\bm{x}_1 - \bm{x}_2)^m\}] \sim \begin{modnew}
        r^D
    \end{modnew}, \\
    E \{\cdots \xi(\bm{x}_1 - \bm{x}_2)\xi(\bm{x}_1 - \bm{x}_3)\} \le E \{\xi(\bm{x}_1 - \bm{x}_2)\xi(\bm{x}_1 - \bm{x}_3)\} = \begin{modnew}
        O(r^D)
    \end{modnew},\\
    E \{\cdots \xi(\bm{x}_1 - \bm{x}_2)\xi(\bm{x}_3 - \bm{x}_4)\} \le E \{\xi(\bm{x}_1 - \bm{x}_2)\xi(\bm{x}_3 - \bm{x}_4)\} = \begin{modnew}
        O(r^D)
    \end{modnew}.
\end{gather*}
Therefore, we only need to keep those $M\{(j_\ell, k_\ell)_{\ell=1}^m\}$ whose sequence $(j_\ell, k_\ell)_{\ell=1}^m$ contains two unique values. We have a total of $2^{m-1}n(n-1)$ such sequences and the corresponding $M\{(j_\ell, k_\ell)_{\ell=1}^m\}$ sum to
\begin{align*}
    2^{m-1}n(n-1)M\left\{(j_\ell, k_\ell)_{\ell=1}^m\right\} = A_m \begin{modnew}
        +O(r^D)
    \end{modnew},
\end{align*}
where $A_m = \{2^{m-1}n(n-1)/n^m\}E [\{\xi(\bm{x}_1 - \bm{x}_2)\}^m]$, $m \ge 2$. When $m = 1$, $M\{(j_1, k_1)\} = 0$ if $j_1 \neq k_1$, so we define $A_1 = 0$.

Now, suppose there are exactly $p$ pairs of $(j_\ell, k_\ell)$ such that $j_\ell = k_\ell$,\; $p = 1, \ldots, m - 1$. For such an index $\ell$, the corresponding factor simplifies to $1 - 2\xi(\bm{x}_{j_\ell} - \bm{y}_\ell) + \xi(\bm{y}_{\ell} - \bm{y}_{\ell}')$. Its last two terms $- 2\xi(\bm{x}_{j_\ell} - \bm{y}_\ell) + \xi(\bm{y}_{\ell} - \bm{y}_{\ell}')$ can be dropped, because combined with the remaining factors for which $j_\ell \neq k_\ell$, the whole product contains more than two unique random vectors and thus is $\begin{modnew}
        O(r^D)
    \end{modnew}$. Therefore, $M\{(j_\ell, k_\ell)_{\ell=1}^m\} = n^{-m}E [\{\xi(\bm{x}_1 - \bm{x}_2)\}^{m-p}] \begin{modnew}
        +O(r^D)
    \end{modnew}$. Since there are a total of $\binom{m}{p}n^p \cdot 2^{m-1-p}n(n-1)$ required sequences, $M\{(j_\ell, k_\ell)_{\ell=1}^m\}$ sum to
\[ \binom{m}{p}n^p 2^{m-1-p}n(n-1) M\left\{(j_\ell, k_\ell)_{\ell=1}^m\right\} = \binom{m}{p}A_{m-p} \begin{modnew}
        +O(r^D)
    \end{modnew}. \]

When every pair $(j_\ell, k_\ell)$ satisfies $j_\ell = k_\ell$, i.e., $p = m$, $M\{(j_\ell, k_\ell)_{\ell=1}^m\} = n^{-m}[1 - mE \{\xi(\bm{x}_1 - \bm{x}_2)\}] \begin{modnew}
        +O(r^D)
    \end{modnew}$. There are $n^m$ such sequences. Combined with the results we have obtained earlier, the $m$th noncentral moment is
\begin{align}
    \label{eq:noncentral_value}
    \mu_m' &= \sum_{\substack{j_\ell, k_\ell = 1\\ (\ell = 1,\ldots,m)}}^n M\left\{(j_\ell, k_\ell)_{\ell=1}^m\right\}\nonumber\\
    &=\sum_{p=0}^{m-1} \binom{m}{p} A_{m-p} + 1 - mE \{\xi(\bm{x}_1 - \bm{x}_2)\} \begin{modnew}
        +O(r^D)
    \end{modnew}. 
\end{align}

Crucially, $\mu_m'$ can also be expressed in a similar way using central moments:
\begin{align*}
    \mu_m' &= E \left\{(\Delta - \mu_1' + \mu_1')^m\right\}\\
    &= \sum_{p = 0}^{m} \binom{m}{p} \mu_1'^p \mu_{m-p}\\
    &= \sum_{p = 0}^{m} \binom{m}{p} \bigl[1 - E \{\xi(\bm{x}_1 - \bm{x}_2)\}\bigr]^p \mu_{m-p}\\
    &= \sum_{p = 0}^{m-2} \binom{m}{p} \bigl[1 - E \{\xi(\bm{x}_1 - \bm{x}_2)\}\bigr]^p \mu_{m-p} + \bigl[1 - E \{\xi(\bm{x}_1 - \bm{x}_2)\}\bigr]^m\\
    &= \sum_{p = 0}^{m-1} \binom{m}{p} \bigl[1 - E \{\xi(\bm{x}_1 - \bm{x}_2)\}\bigr]^p \mu_{m-p} + 1 - mE \{\xi(\bm{x}_1 - \bm{x}_2)\} \begin{modnew}
        +O(r^D)
    \end{modnew}.
\end{align*}
The range of summation can be changed in the last step because $\mu_1 = 0$. Subtracting by Equation \eqref{eq:noncentral_value} gives
\[ \sum_{p = 0}^{m-1} \binom{m}{p} \Bigl[ \bigl[1 - E \{\xi(\bm{x}_1 - \bm{x}_2)\}\bigr]^p \mu_{m-p} - A_{m-p}\Bigr] = \begin{modnew}
        O(r^D)
    \end{modnew}. \]
Using mathematical induction, we can easily show that
\begin{align*}
    \mu_m &= A_{m} \begin{modnew}
        +O(r^D)
    \end{modnew}\\
    &= (n-1)\left(\frac{2}{n}\right)^{m-1}\left(\frac{2}{m}\right)^D\begin{modnew}
        r^D
    \end{modnew} \begin{modnew}
        +O(r^D)
    \end{modnew} \quad (m \ge 2).
\end{align*}

To derive the cumulants, we use a recursive relationship between the central moments and cumulants \citep{Willink2003},
\begin{gather*}
    \kappa_m = \mu_m - \sum_{\ell=1}^{m-2} \binom{m-1}{\ell}\mu_{\ell}\kappa_{m-\ell} \quad (m \ge 2).
\end{gather*}
It can be shown again by mathematical induction that
\begin{align*}
    \kappa_m &= \mu_m \begin{modnew}
        +O(r^D)
    \end{modnew}\\
    &= (n-1)\left(\frac{2}{n}\right)^{m-1}\left(\frac{2}{m}\right)^D\begin{modnew}
        r^D
    \end{modnew}\begin{modnew}
        +O(r^D)
    \end{modnew} \quad (m \ge 2).
\end{align*}

\bibliographystyle{biometrika}
\bibliography{spatial}

\begin{thebibliography}{40}
\expandafter\ifx\csname natexlab\endcsname\relax\def\natexlab#1{#1}\fi

\bibitem[{Arbenz et~al.(1988)Arbenz, Gander \& Golub}]{ARBENZ198875}
\textsc{Arbenz, P.}, \textsc{Gander, W.} \& \textsc{Golub, G.~H.} (1988).
\newblock Restricted rank modification of the symmetric eigenvalue problem: Theoretical considerations.
\newblock \textit{Linear Algebra and its Applications} \textbf{104}, 75--95.

\bibitem[{Baddeley et~al.(2015)Baddeley, Rubak \& Turner}]{baddeley2015spatial}
\textsc{Baddeley, A.}, \textsc{Rubak, E.} \& \textsc{Turner, R.} (2015).
\newblock \textit{Spatial Point Patterns: Methodology and Applications with R}.
\newblock Chapman \& Hall/CRC Interdisciplinary Statistics. CRC Press.

\bibitem[{Baringhaus \& Henze(1988)}]{baringhaus1988}
\textsc{Baringhaus, L.} \& \textsc{Henze, N.} (1988).
\newblock {A consistent test for multivariate normality based on the empirical characteristic function}.
\newblock \textit{Metrika: International Journal for Theoretical and Applied Statistics} \textbf{35}, 339--348.

\bibitem[{Blatter(2014)}]{MathSE}
\textsc{Blatter, C.} (2014).
\newblock The probability that the distance between two random points in a $3\times 3$ square is less than $\sqrt{3}$.
\newblock Mathematics Stack Exchange.
\newblock URL:https://math.stackexchange.com/q/992314 (version: 2014-10-26).

\bibitem[{Bunch et~al.(1978)Bunch, Nielsen \& Sorensen}]{Bunch1978}
\textsc{Bunch, J.~R.}, \textsc{Nielsen, C.~P.} \& \textsc{Sorensen, D.~C.} (1978).
\newblock Rank-one modification of the symmetric eigenproblem.
\newblock \textit{Numerische Mathematik} \textbf{31}, 31--48.

\bibitem[{Clark \& Evans(1954)}]{clarkevans}
\textsc{Clark, P.~J.} \& \textsc{Evans, F.~C.} (1954).
\newblock Distance to nearest neighbor as a measure of spatial relationships in populations.
\newblock \textit{Ecology} \textbf{35}, 445--453.

\bibitem[{Diggle et~al.(1987)Diggle, Gates \& Stibbard}]{diggle1987}
\textsc{Diggle, P.~J.}, \textsc{Gates, D.~J.} \& \textsc{Stibbard, A.} (1987).
\newblock A nonparametric estimator for pairwise-interaction point processes.
\newblock \textit{Biometrika} \textbf{74}, 763--770.

\bibitem[{Donnelly(1978)}]{donnelly1978}
\textsc{Donnelly, K.} (1978).
\newblock Simulations to determine the variance and edge-effect of total nearest neighbour distance.
\newblock In \textit{Simulation Methods in Archaeology}, I.~Hodder, ed. Cambridge University Press, pp. 91--95.

\bibitem[{Epps(2005)}]{epps2005}
\textsc{Epps, T.~W.} (2005).
\newblock Tests for location-scale families based on the empirical characteristic function.
\newblock \textit{Metrika} \textbf{62}, 99--114.

\bibitem[{Epps \& Pulley(1983)}]{epps1983}
\textsc{Epps, T.~W.} \& \textsc{Pulley, L.~B.} (1983).
\newblock A test for normality based on the empirical characteristic function.
\newblock \textit{Biometrika} \textbf{70}, 723--726.

\bibitem[{Eubank \& LaRiccia(1992)}]{eubank1992}
\textsc{Eubank, R.~L.} \& \textsc{LaRiccia, V.~N.} (1992).
\newblock Asymptotic comparison of {C}ramer-von {M}ises and nonparametric function estimation techniques for testing goodness-of-fit.
\newblock \textit{The Annals of Statistics} \textbf{20}, 2071--2086.

\bibitem[{Fan(1997)}]{FAN1997}
\textsc{Fan, Y.} (1997).
\newblock Goodness-of-fit tests for a multivariate distribution by the empirical characteristic function.
\newblock \textit{Journal of Multivariate Analysis} \textbf{62}, 36--63.

\bibitem[{Gerrard(1969)}]{gerrard1969competition}
\textsc{Gerrard, D.~J.} (1969).
\newblock \textit{Competition Quotient: a New Measure of the Competition Affecting Individual Forest Trees}, vol.~20.
\newblock Agricultural Experiment Station, Michigan State University.

\bibitem[{Gil-Pelaez(1951)}]{gil-pelaez}
\textsc{Gil-Pelaez, J.} (1951).
\newblock Note on the inversion theorem.
\newblock \textit{Biometrika} \textbf{38}, 481--482.

\bibitem[{Golub(1973)}]{golub1973}
\textsc{Golub, G.~H.} (1973).
\newblock Some modified matrix eigenvalue problems.
\newblock \textit{SIAM Review} \textbf{15}, 318--334.

\bibitem[{Ho \& Chiu(2007)}]{Ho2007}
\textsc{Ho, L.~P.} \& \textsc{Chiu, S.~N.} (2007).
\newblock Testing uniformity of a spatial point pattern.
\newblock \textit{Journal of Computational and Graphical Statistics} \textbf{16}, 378--398.

\bibitem[{Illian et~al.(2007)Illian, Penttinen, Stoyan \& Stoyan}]{illian2007intro}
\textsc{Illian, J.}, \textsc{Penttinen, A.}, \textsc{Stoyan, H.} \& \textsc{Stoyan, D.} (2007).
\newblock The homogeneous poisson point process.
\newblock In \textit{Statistical Analysis and Modelling of Spatial Point Patterns}, chap.~2. John Wiley \& Sons, Ltd, pp. 57--98.

\bibitem[{Imhof(1961)}]{imhof}
\textsc{Imhof, J.~P.} (1961).
\newblock Computing the distribution of quadratic forms in normal variables.
\newblock \textit{Biometrika} \textbf{48}, 419--426.

\bibitem[{Jiménez-Gamero et~al.(2009)Jiménez-Gamero, Alba-Fernández, Muñoz-García \& Chalco-Cano}]{JIMENEZGAMERO20093957}
\textsc{Jiménez-Gamero, M.~D.}, \textsc{Alba-Fernández, V.}, \textsc{Muñoz-García, J.} \& \textsc{Chalco-Cano, Y.} (2009).
\newblock Goodness-of-fit tests based on empirical characteristic functions.
\newblock \textit{Computational Statistics \& Data Analysis} \textbf{53}, 3957--3971.

\bibitem[{Koutrouvelis(1980)}]{koutrouvelis1980}
\textsc{Koutrouvelis, I.~A.} (1980).
\newblock A goodness-of-fit test of simple hypotheses based on the empirical characteristic function.
\newblock \textit{Biometrika} \textbf{67}, 238--240.

\bibitem[{Koutrouvelis \& Kellermeier(1981)}]{koutrouvelis1981}
\textsc{Koutrouvelis, I.~A.} \& \textsc{Kellermeier, J.} (1981).
\newblock A goodness-of-fit test based on the empirical characteristic function when parameters must be estimated.
\newblock \textit{Journal of the Royal Statistical Society. Series B (Methodological)} \textbf{43}, 173--176.

\bibitem[{Li(1994)}]{li1994}
\textsc{Li, R.-C.} (1994).
\newblock Solving secular equations stably and efficiently.
\newblock Tech. Rep. UCB/CSD-94-851.

\bibitem[{Mat{\'e}rn(1986)}]{Matérn1986}
\textsc{Mat{\'e}rn, B.} (1986).
\newblock Some particular models.
\newblock In \textit{Spatial Variation}, chap.~3. New York, NY: Springer New York, pp. 27--51.

\bibitem[{Meintanis(2016)}]{Meintanis}
\textsc{Meintanis, S.} (2016).
\newblock A review of testing procedures based on the empirical characteristic function.
\newblock \textit{South African Statistical Journal} \textbf{50}, 1--14.

\bibitem[{Møller et~al.(1998)Møller, Syversveen \& Waagepetersen}]{Møller}
\textsc{Møller, J.}, \textsc{Syversveen, A.~R.} \& \textsc{Waagepetersen, R.~P.} (1998).
\newblock Log {G}aussian {C}ox processes.
\newblock \textit{Scandinavian Journal of Statistics} \textbf{25}, 451--482.

\bibitem[{Numata(1961)}]{numata1961}
\textsc{Numata, M.} (1961).
\newblock Forest vegetation in the vicinity of {C}hoshi: coastal flora and vegetation at {C}hoshi, {C}hiba {P}refecture {IV} (in {J}apanese).
\newblock \textit{Bulletin of Choshi Marine Laboratory, Chiba University} \textbf{3}, 28--48.

\bibitem[{P{\'o}lya(1949)}]{polya1949remarks}
\textsc{P{\'o}lya, G.} (1949).
\newblock Remarks on characteristic functions.
\newblock In \textit{Proc. First Berkeley Conf. on Math. Stat. and Prob}. University of California Press, pp. 115--123.

\bibitem[{Rajala et~al.(2023)Rajala, Olhede, Grainger \& Murrell}]{rajala}
\textsc{Rajala, T.}, \textsc{Olhede, S.}, \textsc{Grainger, J.} \& \textsc{Murrell, D.} (2023).
\newblock What is the {F}ourier transform of a spatial point process?
\newblock \textit{IEEE Transactions on Information Theory} \textbf{PP}, 1--1.

\bibitem[{Ripley(1977)}]{Ripley1977}
\textsc{Ripley, B.~D.} (1977).
\newblock Modelling spatial patterns.
\newblock \textit{Journal of the Royal Statistical Society. Series B (Methodological)} \textbf{39}, 172--212.

\bibitem[{Ripley(1978)}]{Ripley1978}
\textsc{Ripley, B.~D.} (1978).
\newblock The analysis of geographical maps.
\newblock In \textit{Exploratory and Explanatory Statistical Analysis of Spatial Data}, C.~P.~A. Bartels \& R.~H. Ketellapper, eds., chap.~3. Dordrecht: Springer Netherlands, pp. 53--72.

\bibitem[{Ripley(1979)}]{ripley1979}
\textsc{Ripley, B.~D.} (1979).
\newblock Tests of `randomness' for spatial point patterns.
\newblock \textit{Journal of the Royal Statistical Society. Series B (Methodological)} \textbf{41}, 368--374.

\bibitem[{Ripley(1988)}]{Ripley1988}
\textsc{Ripley, B.~D.} (1988).
\newblock Edge corrections for spatial point processes.
\newblock In \textit{Statistical Inference for Spatial Processes}, chap.~3. Cambridge University Press, p. 22–48.

\bibitem[{Sato(2004)}]{sato2004remarks}
\textsc{Sato, K.} (2004).
\newblock Remarks on {P}ólya’s theorem on characteristic functions.
\newblock \textit{Institute of Statistical Mathematics, Cooperative Res. Rep.} , 133--145.

\bibitem[{Serfling(1980)}]{serfling1980approximation}
\textsc{Serfling, R.~J.} (1980).
\newblock \textit{Approximation Theorems of Mathematical Statistics}.
\newblock Wiley Series in Probability and Statistics. Wiley, pp. 226--227.

\bibitem[{Shaked \& Shanthikumar(2007)}]{shaked2007stochastic}
\textsc{Shaked, M.} \& \textsc{Shanthikumar, J.~G.} (2007).
\newblock Multivariate stochastic orders.
\newblock In \textit{Stochastic Orders}, Springer Series in Statistics, chap.~6. Springer New York, pp. 265--266.

\bibitem[{Shephard(1991)}]{shephard1991a}
\textsc{Shephard, N.} (1991).
\newblock From characteristic function to distribution function: A simple framework for the theory.
\newblock \textit{Econometric Theory} \textbf{7}, 519--529.

\bibitem[{Strauss(1975)}]{strauss1975}
\textsc{Strauss, D.~J.} (1975).
\newblock A model for clustering.
\newblock \textit{Biometrika} \textbf{62}, 467--475.

\bibitem[{Willink(2003)}]{Willink2003}
\textsc{Willink, R.} (2003).
\newblock Relationships between central moments and cumulants, with formulae for the central moments of gamma distributions.
\newblock \textit{Communications in Statistics - Theory and Methods} \textbf{32}, 701--704.

\bibitem[{Zimmerman(1993)}]{zimmerman1993}
\textsc{Zimmerman, D.~L.} (1993).
\newblock A bivariate {C}ramér–von {M}ises type of test for spatial randomness.
\newblock \textit{Journal of the Royal Statistical Society Series C: Applied Statistics} \textbf{42}, 43--54.

\bibitem[{Zimmerman(1994)}]{ZIMMERMAN1994189}
\textsc{Zimmerman, D.~L.} (1994).
\newblock On the limiting distribution of and critical values for an origin-invariant bivariate {C}ramér–von {M}ises-type statistic.
\newblock \textit{Statistics \& Probability Letters} \textbf{20}, 189--195.

\end{thebibliography}

\end{document}